# Non-Absoluteness, Tracking, and Intersubjectivity in Perspectival Interpretations of Quantum Mechanics: A Conceptual and Formal Analysis

Hervé Zwirn

*Centre Borelli (ENS Paris Saclay, France) & IHPST (CNRS, Université Paris 1)*
*herve.zwirn@gmail.com*

**Abstract:** This paper offers a conceptual and formal analysis of non-absoluteness in perspectival interpretations of quantum mechanics. Extended Wigner-type experiments reveal a fundamental tension between the universality of quantum mechanics, the absoluteness of observed events, and the possibility of intersubjective agreement. I argue that denying observer-independent events cannot remain a mere interpretive slogan: it requires a precise account of how perspectival statements, tracking assumptions, meta-level descriptions, and intersubjective communication are to be formulated. Many existing approaches either implicitly reintroduce observer-independent structure or leave key notions underdetermined. Within this framework, Convivial Solipsism provides a coherent account of non-absolute events, preserving the universality of quantum mechanics while rejecting the need for a global viewpoint. To make this claim precise, the paper introduces the basic elements of a formal perspectival logic designed to distinguish well-formed perspectival statements from illegitimate unindexed or absolutizing statements. The paper concludes that scientific objectivity need not rest on observer-independent facts, but can instead be grounded in the internal coherence of perspectival structures.

## 1. Introduction

Quantum mechanics (QM) provides exceptionally well-confirmed predictions, yet there is no generally agreed interpretation of what it tells us about the world. Competing interpretations differ in their accounts of reality, the quantum state, and measurement outcomes. These differences become particularly acute in Wigner-type scenarios, where the status of outcomes and their comparability across observers are directly challenged.

A central difficulty in this context is that the conclusion often drawn from extended Wigner-type analyses—namely that observed events are not absolute—remains insufficiently specified. Simply denying the absoluteness of events does not by itself provide a coherent interpretive framework. Without a precise account, this notion risks either remaining ambiguous or implicitly reintroducing a form of observer-independent factuality through hidden assumptions about the comparability of outcomes.

The aim of this paper is not to prove a new no-go theorem, but to clarify the conceptual grammar required by interpretations that reject the absoluteness of observed events. Its contribution is to show that debates about extended Wigner's friend scenarios depend crucially on often implicit assumptions about tracking, meta-level description, and the comparison of perspectives. Non-absoluteness must be given a rigorous conceptual structure. This requires clarifying how statements about different observers can be formulated, what role meta-level descriptions play, and under which conditions comparisons between perspectives are meaningful. From this point of view, extended Wigner-type scenarios are not only no-go results, but tools for identifying the structural constraints that any coherent interpretation must satisfy.

In this respect, the paper contributes to the current philosophical debate on observer-independent facts, relational or perspectival interpretations, and the conditions under which scientific objectivity can survive the rejection of absolute events. The contribution of this paper is threefold. First, it identifies Tracking as the hidden assumption that often allows non-absolute events to be implicitly re-absolutized. Second, it shows that intersubjectivity need not be grounded in observer-independent facts, but can be understood as internal coherence within perspectival structures. Third, it introduces the basic elements of a formal perspectival logic designed to distinguish legitimate indexed statements from illegitimate unindexed or absolutizing claims.

Section 2 presents the measurement problem and introduces a taxonomy of interpretations. Section 3 discusses the Wigner's friend experiment. Section 4 analyses the concept of universality of quantum mechanics[1]. Section 5 introduces the so-called Extended Wigner's Friend Experiments (EWF) and the associated no-go theorems. Some issues with the hypotheses used in these no-go theorems are analysed in section 6. In section 7, we will examine how Convivial Solipsism (ConSol) responds to these no-go theorems. In particular, we will give a precise description of what a measurement is in ConSol and how that provides answers to the questions left open in the previous sections. The appendix introduces the associated formal language.

In the following, we will rely on the excellent review papers (Leifer, 2014; Schmid et al., 2023, 2025) and we will draw extensively on them without citing them again each time.

# 2. The measurement problem

## 2.1 Description

As is well known, the measurement problem stems from the existence of two different laws of evolution that lead to results that cannot be reconciled. The first of these laws (the Schrödinger equation) applies when the system is not subject to any measurements. The second (the projection postulate) applies when the system is measured. This would be perfectly acceptable if it were possible to give a clear and unambiguous definition of what a measurement is. But the quantum formalism is unable to do so.

The measurement problem is expressed very simply in the formalism of the standard interpretation, the so-called orthodox interpretation[2] which is that of the textbooks and which contains the postulate of projection. This postulate states that after a measurement of an observable which has yielded as a result one of the possible eigenvalues, the state vector of the system must be projected onto the eigenvector associated with this eigenvalue (or onto the corresponding sub-eigenspace in case of a degenerate eigenvalue). We can give a description of the problem within the framework of the orthodox interpretation using the example of the measurement of a qubit.

We consider a qubit S prepared in the state $|+\rangle_S = \frac{1}{\sqrt{2}}\{|0\rangle_S + |1\rangle_S\}$ on which we make a measurement in the basis $\{|0\rangle_S, |1\rangle_S\}$. The quantum formalism tells us that we can obtain the result $|0\rangle_S$ or the result $|1\rangle_S$ each with probability ½. Moreover, the projection postulate tells us that after the measurement, the state of the system is no longer $|+\rangle_S$ but must be changed to $|0\rangle_S$ in the first case and to $|1\rangle_S$ in the second.

This measurement is made using an apparatus M that interacts with the system S during the measurement time. This interaction is governed by an interaction Hamiltonian $H_{SM}$. The evolution of the ensemble (system plus device) is given by the Schrödinger equation applied to the state vector of the ensemble. The law of evolution given by the Schrödinger equation can be represented by the action of a unitary transformation $U_{SM}$ on the state vector of the ensemble. Let us assume that before the measurement, the state vector of the device is $|R\rangle_M$ (as is customary to indicate that the device is ready to make the measurement). In this case, the fact that the device is working correctly means that if we have prepared the system in state $|0\rangle_S$ or in the state $|1\rangle_S$ we should have after the measurement:

$U_{SM}|0\rangle_S|R\rangle_M = |0\rangle_S|0\rangle_M$ (where $|0\rangle_M$ is the state of the apparatus showing the result 0)

$U_{SM}|1\rangle_S|R\rangle_M = |1\rangle_S|1\rangle_M$ (where $|1\rangle_M$ is the state of the apparatus showing the result 1).

Hence, the transformation $U_{SM}$ applied to the initial state vector when the system is in state $|+\rangle_S$ gives:

$$U_{SM}\,(|+\rangle_S|R\rangle_M) = \frac{1}{\sqrt{2}}\{|0\rangle_S|0\rangle_M + |1\rangle_S|1\rangle_M\}$$

We can therefore see that depending on whether we apply the projection postulate or the unitary evolution to the system, we end up with two different results. In the first case, after the measurement, the system is assumed to be in one of the two states $|0\rangle_S$ or $|1\rangle_S$. In the second case, the system is in an entangled state with the apparatus that does not correspond to any defined value of the measured observable. There is therefore a contradiction which is rarely analysed in more detail. So let us take a closer look at where the contradiction comes from.

The first thing is that the formalism we use contains both the projection postulate and the unitary evolution postulate, and that both are applied to the same situation, for which there is nothing to specify which of the two is to be used to the exclusion of the other. It goes without saying, then, that the problem cannot arise in formalisms that take into account only one of the two rules of evolution. This is the case with spontaneous collapse theories (GRW or Diosi-Penrose) (Diósi, 1989; Ghirardi et al., 1986; Penrose, 1996), in which evolution is not unitary, or theories that do not have a projection postulate (Everett interpretation [DeWitt & Graham, 1973; Everett, 1957a, 1957b], de Broglie-Bohm theory) (Bohm, 1952a, 1952b). We will nevertheless exclude these theories from our analysis either because they modify the standard formalism or because they accept multiple outcomes.

Let us now ask ourselves why it is awkward to end up with two possibly different state vectors for the same system after a measurement. If we assume that the state vector is general, i.e. that all observers must necessarily have the same one, that it is in a one-to-one correspondence with the real physical state of the system, and that it describes a physical reality that is shared by all observers, then it is clear that the same system cannot be associated with two different state vectors, and therefore that at least one of them is wrong. The orthodox interpretation which gives no way to choose between the two is therefore either incomplete or contradictory. Note that our reasoning makes no explicit reference to any observer. The measurement experiment is described in a neutral way, as if this description were valid for any observer whatsoever[3].

[1] What is called the universality of quantum mechanics in these debates is the fact that it is possible to apply the unitary formalism (without projection postulate) to all systems, including macroscopic ones and observers.

[2] Some authors use 'orthodox interpretations' to refer to the relativist interpretations considered below. But we prefer to call orthodox the usual and most widely used interpretation which is the one written in most student textbooks, sometimes called TBQM for Textbook Quantum Mechanics.

The assumed hypotheses are:

- H1: There is a physical reality independent of observers and shared by them. In this reality, systems have their own physical state (not necessarily associated with precise values for all properties).
- H2: The state vector of the quantum formalism associated with a system is identical for all observers.
- H3: There is a one-to-one correspondence between the state vector and the physical state of the system.
- H4: The evolution of a system is described by a unitary interaction given by Schrödinger's equation, and the projection postulate applies when a result has been obtained from a measurement.

Assumption H3 is associated with what is often called the eigenvector-eigenvalue link, which means that the system possesses a certain property when it is in the corresponding eigenstate of the relevant observable, and only then.

Hypotheses H1 to H4 jointly lead to a contradiction, but we can see that there are several ways of escaping it. We can, of course, reject the projection postulate, as we saw above in the interpretations with a modified formalism we cited. Here we are interested in what can be done within the framework of an unmodified formalism (although not necessarily incorporating the projection postulate). Firstly, we can reject H1, i.e. deny that there is a reality shared by all observers (either that there is no reality, or that it is relative to each observer). Secondly, we can accept the existence of a shared reality, but consider that different observers may not have the same state vector (thus denying H2). In this case, the state vector can obviously no longer represent reality one-to-one. Thirdly, we can accept that there is a reality shared by all observers and that the state vector is the same for all observers, but renounce the one-to-one correspondence between the state vector and the physical state of the system (denying H3). Alternatively, we can accept only a watered-down version of H4, assuming that unitary evolution does not apply to all systems (in particular that it does not apply to macroscopic systems). These different possibilities are interrelated and may, depending on how they are realised, imply each other.

In what follows, we will attempt to build a taxonomy, and see how the various interpretations relate to these possibilities. But let us begin by analysing an example of an interpretation that is often claimed to escape contradiction. The analysis proposed here is not intended to be exhaustive, but rather to serve as an example of the reasons that will lead us later on to propose a list of questions to which it is desirable to have answers prior to any reasoning concerning a given experimental situation.

## 2.2 Is it enough to say that the state vector does not represent a physical state?

Some interpretations do not consider that the state vector directly represents the physical state of the system, but rather the observer's knowledge of it. These interpretations are referred to as ψ-epistemic according to a taxonomy we will rely on later, which is due to Harrigan and Spekkens (Harrigan & Spekkens, 2010; Spekkens, 2007). They modify hypothesis H4 to state that the unitary evolution is not that of the physical state of the system, but that of the observer's knowledge, while retaining the postulate of reduction, which takes on the meaning of updating information after taking cognizance of the measurement result. They may or may not accept hypothesis H1, but they reject hypotheses H2 and H3. It is then explained that the two different state vectors we contrasted above are not in contradiction, but simply represent state vectors for observers who do not have the same information about the system. The reduced vector (e.g. $|0\rangle_S$) could be that adopted by the observer who made the measurement and obtained the result 0, while the vector $\frac{1}{\sqrt{2}}\{|0\rangle_S|0\rangle_M + |1\rangle_S|1\rangle_M\}$ would be that of an observer who did not witness the measurement and just knows that there was an interaction between the system and the device.

But this way out is not entirely satisfactory, for two reasons. First of all, we are entering into a discussion that we will go into in greater depth in the case of Wigner's friend experiment, but we can already say that the observer, not knowing the result but knowing that a measurement has taken place, should rather adopt a description corresponding to a classical uncertainty if they admit that the interaction between the measuring apparatus and the system has produced a definite result. In this case, they should not assign an entangled state to the composite system but rather a density operator giving a probability ½ of having obtained $|0\rangle_S|0\rangle_M$ and ½ of having obtained $|1\rangle_S|1\rangle_M$. Of course, it is possible here (where there are only a system and an apparatus) to object that the simple interaction with the measuring device has not produced a definite result and that, in the absence of any additional information, the observer can legitimately use the entangled state vector. We shall see later in the case of Wigner's friend that this is no longer possible, and we shall then question the legitimacy of the choice between the entangled state vector and the density operator expressing a classical uncertainty. But a second reason for dissatisfaction can be put forward, although it is not a real contradiction.

[3] Even more, it is described as if there were no need for any observer in a measurement.

This reason, developed more extensively in a previous paper (Zwirn, 2024), could be expressed by the question “Is it enough to say that the state vector does not represent a physical state to solve the measurement problem?” This issue becomes particularly acute in Wigner-type scenarios.

First, if the reduction of the wave function is interpreted as a mere update of knowledge after the measurement, this raises the question “knowledge about what?” (see Zwirn, 2023 for a discussion of this point). The Kochen-Specker theorem (Kochen & Specker, 1967) shows that, in Hilbert spaces of dimension at least three, it is impossible to assign definite, context-independent values simultaneously to all relevant observables while respecting the functional relations required by quantum mechanics. As a result, the naïve realist view, in which an observer can simply passively observe these properties (hence can use a classical probability calculation that merely reflects their ignorance), must be abandoned in favour of a view in which the system reveals its properties only when experiments are directed towards some of them to the exclusion of others, and reveals them in a non-deterministic way. In classical physics, the observation of an event is independent of the occurrence of the event. This is not the case in quantum mechanics. Then even if we must recognise that the conflict between the unitary evolution of the state vector and the reduction postulate (which becomes a simple updating of our knowledge) has disappeared, it remains to explain what happened that created the values that the observer sees. In other words, what is the ontology of the interpretation? This is a question that is largely neglected in the analysis of this type of interpretation. QBists typically leave this question open (Fuchs & Schack, 2013, 2014, 2015; Fuchs & Stacey, 2019) as explained in Zwirn (2022). They say that the world has no laws but simultaneously contend that the agent’s experience creates the result, but the mechanism remains unspecified. Pragmatists often say that getting a definite result is an empirical fact that is outside the scope of physics. But all this leaves the ontology underdetermined… Worse, they sometimes make controversial use of decoherence to claim that it provides a definite result even when no observer is there, which is a misunderstanding of how decoherence works (Healey, 2012, 2016, 2017, 2025).

The pragmatist attitude is analogous to calculating the probabilities of a die’s outcomes while excluding the mechanics of its motion from scientific explanation. In my view, an interpretation of quantum mechanics should likewise address how definite outcomes arise.

## 2.3 A series of questions

Rigorously discussing what happens in the various experimental situations examined therefore requires specifying the interpretive framework being used. Here are some questions to which it seems necessary to have answers; otherwise several lines of reasoning that analyse them rely on concepts that are insufficiently explained and this can lead to ambiguous statements or even errors in reasoning. They are not meant to be exhaustively answered at this stage, but to structure the conceptual landscape:

- Regarding reality (questions R):
  - R1: Does reality exist?
  - R2: If reality exists, is it the same for everyone?
  - R3: Is this reality (whether shared or individual) comprehensible?
  - R4: If it is comprehensible, is quantum mechanics a means of describing it?
  - R5: If yes to R4, does quantum mechanics describe reality completely or partially?
  - R6: If yes to R4, is quantum mechanics universal, i.e. does it apply to all systems, including macroscopic systems and observers?

- Regarding systems (questions S):
  - S1: Do systems have their own real physical state outside of any measurement and independently of any observer?
  - S2: If so, is this real state knowable by measurement or is it fundamentally beyond reach?
  - S3: Do systems possess properties on their own, or only in relation to a measuring device (a context)?

- Regarding the state vector (questions V):
  - V1: Does the state vector represent the physical state of the system?
  - V2: If yes, is it a complete or partial representation?
  - V3: If not, does it only represent the information the observer has about the system?
  - V4: Is it merely a tool to calculate the probability of the next outcome?
  - V5: Is the state vector the same for all observers, or is it relative to each observer?

- Regarding measurement (questions M):
  - M1: Is measurement a purely physical process, i.e., can a measurement occur through the interaction between a system and an apparatus alone (in the absence of any observer)?
  - M2: If yes, and since it is generally not possible to assume that the values obtained during a measurement are pre-existing, can and should the quantum formalism describe the measurement process, i.e. the selection of a single definite outcome?
  - M3: Does this occur independently of any observer as an autonomous phenomenon (in which case the observer merely passively becomes aware of the result), or does the observer play a role in the measurement?
  - M4: If the process leading to the selection of a definite state lies outside the scope of quantum mechanics, should we settle for the empirical observation that during a measurement, a result is indeed obtained, without seeking further explanation?
  - M5: If we do not settle for that, where does the change of the state of the system come from?

- M6: Does a measurement provide a result that is valid for all observers, or a result relative to the observer who performed it?

- Regarding intersubjectivity (question I):

- I1: Is intersubjectivity grounded?

Clearly, the difficulties outlined above will not be of the same nature, depending on the answers given. These questions are obviously not totally independent of one another, but there is still considerable freedom in the choice of answers. A satisfactory interpretation must, at the very least, provide a clear answer to all these questions, while escaping the no-go theorems we will present in the following. However, it is not always possible to give a simple yes or no answer. Sometimes more nuanced answers are given, depending on the interpretation.

In what follows, we will focus on a restricted subset of the most structurally relevant questions and we will be interested only in interpretations that do not modify the formalism so we will not mention GRW or dBB interpretations. We will examine the main interpretations: Everett interpretation (DeWitt & Graham, 1973; Everett, 1957a, 1957b), QBism (Fuchs & Schack, 2013, 2014, 2015; Fuchs & Stacey, 2019), Pragmatism (Healey's version) (Healey, 2012, 2016, 2017, 2025), Contexts, Systems and Modalities (CSM) (Auffèves & Grangier, 2016; Van Den Bossche & Grangier, 2023), the Copenhagen interpretation (Faye, 2019), the orthodox interpretation presented in textbooks (Cohen-Tannoudji et al., 2019), Relational Quantum Mechanics (RQM) (Di Biagio & Rovelli, 2021, 2025; Laudisa & Rovelli, 2021; Rovelli, 1996), the modal interpretation (Dieks's version) (Bene & Dieks, 2002; Dieks, 2022, 2025) and Convivial Solipsism (ConSol) (Zwirn, 2015, 2016, 2020, 2021, 2022, 2023, 2024, 2025, 2026).

Of course, giving a short answer to each question for these interpretations is sometimes not easy because that would require a more detailed discussion but, when possible, we will give what seems to be the closest answer.

| **Question** | **Everett** | **QBism** | **RQM** | **Pragmatism (Healey)** | **CSM** | **TBQM** | **Copenhagen** | **Modal (Dieks)** | **ConSol** |
|---|---|---|---|---|---|---|---|---|---|
| Observer-independent reality? | Yes (universal wavefunction) | No (agent-dependent) | No (relational) | Yes (but epistemically constrained) | Yes (contextual) | Yes | Ambiguous / contextual | Yes | No (perspectives) |
| Single shared reality? | Yes (branch-relative structure) | No | No | Yes (partially inaccessible) | Yes (within context) | Yes | Ambiguous | Yes | No |
| Systems have intrinsic states? | Yes | No | No (only relational properties) | Yes (context-dependent) | No (state tied to system–context pair) | Yes | Often assumed | Yes | No |
| State vector represents reality? | Yes | No (degrees of belief) | No (relational tool) | No (predictive tool) | No (contextual descriptor) | Yes | Ambiguous | Partial | No (interaction state only) |
| Same state for all observers? | Yes | No | No | No | Yes (given same context) | Yes | Often assumed | Yes | No |
| Measurement = physical process? | Yes (unitary) | No (agent action) | Yes (interaction) | Yes (contextual interaction) | Yes (system–context interaction) | Yes | Ambiguous | Yes | No (perspectival) |
| Single outcome? | No (branching) | Yes | Yes (relative) | Yes | Yes | Yes | Yes | Yes | Yes (perspectival) |
| Outcome observer-independent? | No | No | No | Unclear | Yes (within a context) | Yes | Often assumed | Perspective-relative | No |
| Intersubjectivity grounded? | Problematic (branch-relative) | Agent-relative | Relational consistency | Pragmatic agreement | Contextual objectivity | Yes | Implicit | Assumed | Emergent (coherence of perspective) |

**Table 1**

## 2.4 Attempt at a taxonomy

This comparison highlights a key structural difference between interpretations. Many approaches rely on implicit assumptions or leave crucial questions underdetermined, particularly regarding the status of reality, the role of the state vector, and the nature of intersubjectivity.

The table above provides a structured basis for a taxonomy. The two main criteria that will enable us to define this taxonomy are, first, whether or not a realist framework is adopted, and second, the role assigned to the state vector. To do this, we will draw on the classification provided by Harrigan and Spekkens (2010) and another one proposed by Schmid et al. (2025). Nuances can then be added. Harrigan and Spekkens proposed a taxonomy of ontological models that assume an underlying physical state of the system and distinguish different possible relationships between that state and the quantum state. Of course, these interpretations consider that there is a reality independent of observers, and it is implicit that this reality is the same for everyone. Furthermore, quantum mechanics is supposed to describe the state of systems and their evolution. In these interpretations, the system is assumed to have an independent physical state that is neither necessarily directly accessible nor necessarily in one-to-one correspondence with the state vector assigned to it. This is the general framework of the hidden variable theory program. Interpretations in which the state vector represents the actual physical state are called ψ-ontic. On the other hand, if the actual state of the system is compatible with several state vectors, the interpretation will be called ψ-epistemic. The state vector alone is not sufficient to determine the physical state of the system. Rather, it is a representation of the observer's incomplete knowledge of the system. A ψ-epistemic interpretation can be realist, as shown by the (admittedly partial) example of the toy model developed by Spekkens (2007). However, the term ψ-epistemic also extends to non-realist interpretations, as we will see below.

The adjective "realist" can be confusing because it applies to two different concepts. The first is metaphysical realism, which assumes that there is a reality independent of any observer and that would be the same in the absence of observers. This thesis can be denied to varying degrees. Radically, idealism considers that there is no independent reality and that the world is created by the mind. None of the interpretations presented in the table above adopts this position. On the other hand, it is possible to consider that the reality we perceive, although it has an independent part, is partially created or modified by the observer. The second notion of realism concerns the fact that we first consider that the system does indeed have a real physical state and, furthermore, that the state vector is linked to it (either representing it directly or being an incomplete description of it).

Furthermore, the state vector can be considered a complete description of reality, known as the ψ-complete view, or, conversely, it must be associated with additional variables (usually denoted λ), known as the ψ-incomplete view. The additional λ variables are what are commonly referred to as hidden variables because they are not part of standard quantum formalism. Realist ψ-epistemic interpretations are necessarily incomplete. But a ψ-ontic interpretation can be complete or incomplete. The orthodox interpretation in textbooks is ψ-ontic complete, as is Everett's interpretation, albeit in a different way. The Copenhagen Interpretation and CSM are realist but the state vector is not enough to describe the real state of the system. A context is needed as often claimed by Bohr and as emphasized in CSM.

In interpretations that postulate a genuine collapse of an ontic quantum state, collapse is a physical process. Other ψ-ontic interpretations, including Everettian and Bohmian approaches, do not postulate such a collapse. In epistemic approaches, state updating need not represent a physical collapse .

The interpretations that are not realist (in the second meaning that we used above) cannot, of course, be ontic, since the actual physical state of the system is not supposed to exist. They can therefore be described as epistemic, not in the sense that the state vector is compatible with several real physical states, but in the sense that it represents only the knowledge (or belief, or probability guide, or set of recommendations, depending on the interpretation) that the observer has about the system. It represents the observer's knowledge of the system's possible response to an action that they can perform on it. The state vector is therefore no longer a shared object but becomes relative to an observer. In this context, Schmid et al. (2025) proposed calling “Copenhagenish” the interpretations that satisfy the following properties:

1. The quantum state is epistemic (information, knowledge, beliefs, pieces of advice). It is not a direct representation of reality.
2. Outcomes are unique for a given observer.
3. Quantum theory is universal. It applies to every system whatever its size. In particular, observers can be described by quantum theory.
4. Quantum theory is complete. It does not need to be supplemented by hidden variables. There are no fundamental ontic states assigned to systems.

QBism, Convivial Solipsism, Pragmatism, RQM, Brukner’s position (Brukner, 2017, 2018) are Copenhagenish. We can therefore see how the answers to our questions define the boundaries between different types of interpretation. In the following, we will focus on the Copenhagenish interpretations because Everett assumes a multiple-outcomes-per-observer postulate which we do not find acceptable and the others directly face the measurement problem.

# 3. The original Wigner’s friend experiment

The standard presentation of the initial Wigner’s friend experiment (Wigner, 1962) is closely parallel to the presentation we have given of the measurement problem. The only difference is the scenario where a friend of Wigner is supposed to be inside a sealed laboratory while Wigner stays outside. The friend makes a measurement in the basis $\{|0\rangle_S, |1\rangle_S\}$ on a qubit S prepared in the state $|+\rangle_S = \frac{1}{\sqrt{2}}\{|0\rangle_S + |1\rangle_S\}$. He can obtain the result 0 or the result 1 each with probability ½. The projection postulate states that after the measurement, the state of the system is no longer $|+\rangle_S$ but must be updated to $|0\rangle_S$ in the first case and to $|1\rangle_S$ in the second. Simultaneously, the state of the friend will be $|0\rangle_F$ in the first case (the state of the friend having seen 0) and $|1\rangle_F$ in the second (the state of the friend having seen 1). Wigner, for his part, considers the system and the friend as a global system to which he applies a unitary evolution that leads him to adopt the entangled state $|\phi^+\rangle_{SF} = \frac{1}{\sqrt{2}}\{|0\rangle_S|0\rangle_F + |1\rangle_S|1\rangle_F\}$. As in the measurement problem we see that the state attributed to the system by the friend is not the same as the state attributed by Wigner. Moreover, for Wigner, the formalism does not attribute a definite result to the friend since they are neither in the state $|0\rangle_F$ nor in the state $|1\rangle_F$. This situation seems to lead to a contradiction if one assumes that both descriptions refer to a single observer-independent reality. The friend describes a definite outcome, whereas Wigner describes a superposed state. As will become clear below, this apparent contradiction only arises if one assumes that the two descriptions can be combined into a single observer-independent account.

Now it is true that for any series of measurements on the system or on his friend that Wigner can perform, the probabilities given by the state $|\phi^+\rangle_{SF}$ will be in agreement with what he will observe. This is what Adlam (2024) calls a type II disaccord: A state $\psi$ gives correct predictions for the relative frequencies over a large number of trials in all the measurements that Wigner could possibly perform on his friend. But, when intuitively interpreted as representing something about the friend's experience, the "natural" representation of the friend's experience suggested by the state $\psi$ may not agree with what the friend is actually experiencing.

The origin of this tension lies in the assumption of the universality of quantum mechanics (that we will analyse in more detail below), according to which the formalism applies equally to all physical systems, including observers. Wigner's friend scenario thus leads to a type II disaccord. This directly reflects the incompatibility between hypotheses H1–H4 identified in Section 2. This already suggests that the tension does not arise from the formalism itself, but from the assumption that different observational perspectives can be combined into a single observer-independent description.

Let us first recall Wigner's motivation for proposing this thought experiment. The title of the paper (Wigner, 1962) in which Wigner first set out his analysis of this situation is "Remarks on the Mind-Body Question". Its aim is to explore the link between matter and consciousness or mind. After recalling Descartes, who posits mind as a primitive entity, he explicitly adopts the idea that consciousness is the ultimate reality:

> *It will remain remarkable, in whatever way our future concept may develop, that the very study of the external world led to the conclusion that the content of the consciousness is an ultimate reality.*

He insists that quantum mechanics, through the wave function, only provides us with a tool for calculating the probabilities between successive "impressions" that a system produces if we repeatedly interact with it.

He then points out that after an observation has produced a result, we are led to modify the wave function:

> *It is the entering of an impression into our consciousness which alters the wave function because it modifies our appraisal of the probabilities for different impressions which we expect to receive in the future. It is at this point that the consciousness enters the theory unavoidably and unalterably.*

There are two ways of looking at this. The first is the epistemic view. In this view, the wave function does not represent the physical state of the system, but rather the information available to an observer. The update of the state following a measurement is interpreted as a mere update of knowledge and does not represent a physical action on the system. The wave function is then just a tool that describes the probabilities of the results that will be obtained subsequently when the next measurements are made on the system.

In the epistemic account considered here, the measurement interaction is assumed to produce a physical change yielding a definite outcome in a probabilistic manner[4]. The subsequent update of the wave function represents the observer's acquisition of information about that outcome. Interpreting this update as epistemic does not, however, explain the physical process by which the outcome was produced. This epistemic view, in which the wave function represents information that can be communicated between two observers, seems to follow from Wigner's analysis at the beginning of his article:

> *The information given by the wave function is communicable. If someone else somehow determines the wave function of a system, he can tell me about it and, according to the theory, the probabilities for the possible different impressions (or sensations) will be equally large, no matter whether he or I interact with the system in a given fashion.*

[4] The apparent contradiction can be avoided but we must insist, as we did before, that it does not solve the problem of measurement. In fact, the problem is relegated to having to explain how the interaction between the system and the measuring device can lead to a defined result. If we accept the existence of hidden variables such that the state of the system is always defined, then the device will reveal the value due to the hidden variable corresponding to the state of the system. But if we refuse the existence of hidden variables, then we have to admit that the state of the system is really not defined. In this case, explaining how interaction with a device can determine it is not possible within the standard quantum formalism.

Then, the role of consciousness is limited to acquiring information. The system evolves independently of our knowledge of it, and the wave function only describes our knowledge of its state. Updating the wave function reflects the information acquired through the measurement; whether and how the physical state changes is a separate interpretive question.

In this case, there is a very simple way to describe the situation within orthodox quantum mechanics. It suffices to say that the friend uses the projection postulate after having obtained a result coming from the interaction between the system and the apparatus which leads the system to enter into a definite state. Wigner, not knowing the result, but knowing that a measurement has been made, must adopt a proper mixture of the two states of the system. Of course, this supposes that the interaction between the system and the apparatus is by itself sufficient to break the unitarity. So the measurement problem is still there: how can we explain this break in unitarity?

But this relatively innocuous view of consciousness is not the one adopted by Wigner, who makes it clear in what follows that he adopts a second way of looking at things. He asks what happens to the wave function if it is not he who makes the observation, but a friend. His answer is that, in this case, the system itself no longer has a wave function, and that it is the whole, consisting of the system plus his friend, that then possesses a wave function. He then points out that only when he has asked his friend which result he has obtained will the system again have a wave function, corresponding to the result stated by his friend.

On the realist epistemic view considered above, the wave function describes the (transmissible) knowledge of an observer, and another observer may or may not know it, but this in no way affects the state of the first observer, who will not become entangled with the system he is measuring. The analysis that Wigner would have to make in this view would be to say that the friend has made a measurement and obtained information that has led him to assign a new wave function to the system, which does indeed have one. There's nothing mysterious about the fact that Wigner doesn't know it and will only know it by asking his friend. The wave function is an objective, shareable piece of information that some may know and others may not at a given moment. The epistemic realist view therefore eliminates any problem in this situation, but at the cost of having to accept that observers are outside the description of quantum mechanics which is not universal.

In this case, all that remains to be done is to define what an observer is, and to explain how an interaction with a device will precipitate the system into a given state (if we reject the existence of hidden variables). This approach is roughly similar to the Copenhagen interpretation, which separates the measuring apparatus, which (when it is used to perform a measurement) is not described by quantum mechanics, from the system being measured, which is. After interaction with the device, the state of the system is probabilistically projected onto a state such that the value of the measured property is well defined. This is the wave packet projection postulate.

Since it is undeniable that at the end of a measurement process, we do indeed perceive a unique result, at some point unitarity must be broken[5]. As Okon (2025) says:

> *Therefore, on pain of inconsistency, it is simply impossible to assume unitarity and objective outcome, and to employ the standard framework to make predictions.*

This implies that quantum mechanics (excluding the projection postulate) has limited validity. There are then two possibilities: unitarity is broken either when interacting with a macroscopic device or when interacting with a conscious observer. Wigner proposes both solutions and points out that he himself prefers to consider quantum mechanics valid for any inanimate object (and therefore also for macroscopic devices), since the boundary between microscopic and macroscopic is blurred. From this he deduces that it is the fact that the observer is conscious that places him outside the field of quantum mechanics. But by ruling out the possibility that the apparatus plays the role (not described by quantum mechanics) of reducing the wave packet, he cannot take refuge in the relatively benign epistemic position we described above. He is then logically led to accept that it is the observer's consciousness that causes the reduction:

> *It follows that the being with a consciousness must have a different role in quantum mechanics than the inanimate measuring device: the atom considered above. […] The preceding argument for the difference in the roles of inanimate observation tools and observers with a consciousness – hence for a violation of the physical laws where consciousness plays a role – is entirely cogent so long as one accepts the tenets of orthodox quantum mechanics in all their consequences.*

This position is rejected by modern physicists who prefer either to assume that the projection postulate applies FAPP and leave the measurement problem unresolved or, more interestingly, to think that quantum mechanics is universal.

So, let us come back to the conclusion we gave at the end of the presentation of the experiment: "for Wigner, the formalism does not attribute a definite result to the friend since they are neither in the state $|0\rangle_F$ nor in the state $|1\rangle_F$". This conclusion comes from the eigenvalue-eigenstate link which states that a system has a definite value for a property if and only if it is in an eigenstate associated with the observed eigenvalue of the corresponding observable. If we abandon this link, we allow a system to have a definite property even if the state that is attributed to it is not an eigenstate. This is what is possible in epistemic interpretations since the state no longer represents the actual physical state of the system but the knowledge that the observer has. Then it is not problematic that the state attributed by Wigner to his friend be $|\phi^+\rangle_{SF}$ (i.e. neither $|0\rangle_F$ nor $|1\rangle_F$) even if the friend has seen a result.

[5] This is necessary from the orthodox point of view which was the position adopted by Wigner. We will see below that it is no longer true in different views.

To solve the apparent contradiction, Adlam (2024) rightly introduces a distinction between intrinsic states (supposed to refer to the actual physical state of the system) and interaction states (supposed to encode predictions about the results of measurements made on the system by a given observer in a certain context). This is a distinction that I already introduced in previous papers (though without giving a name to the two sides of the state)[6]. In her view, it is the conflation of these two meanings when considering the state of a system that produces the contradiction in Type II disaccords. It is true that, when we consider that the friend has made a measurement inside the laboratory and obtained a result, while Wigner attributes an entangled state to him, it is the fact of interpreting the entangled state as meaning that the friend is in a state where they have not obtained a definite result that causes the contradiction. But if we consider that the entangled state assigned by Wigner is merely an interaction state that does not represent "the real state the friend is in", but is merely a tool for predicting the probabilities of outcomes for the various measurements Wigner may make on his friend, the contradiction disappears. We must accept that from the friend's perspective, a definite outcome has been obtained, while from Wigner's perspective, only probabilistic predictions can be made about the friend's response.

This distinction removes the immediate contradiction. However, it relies on a separation between intrinsic and interaction states whose precise status remains unclear. In particular, it raises the question of whether intrinsic states can be meaningfully defined independently of any observational context.

If we accept the universality of quantum mechanics then Wigner is well justified in attributing the entangled state vector $|\phi^+\rangle_{SF}$ to the composite system {system + friend inside the lab} even though the friend has obtained a result. Wigner can perform experiments on his friend in the same basis as that of the measurement made by the friend. For example, he can ask his friend for the result they got. In a series of experiments of this kind, there will be no statistical discrepancy between the results obtained by Wigner using $|\phi^+\rangle_{SF}$ and the results obtained by the friend. And actually, for all practical purposes, the predictions would be the same using the entangled state or the mixed state in any experiment we can do today inside the box on the system alone or on the friend alone. The disaccord would arise only if Wigner performed a measurement on the whole (system + friend) of the observable $2|\phi^+\rangle_{SF}\langle\phi^+|_{SF} - I$. If universality is true then Wigner should always get the result +1. If universality does not hold and collapse has occurred when the friend got a result, Wigner should get +1 and -1 with probability ½. So, universality is something that is in principle empirically testable even though the kind of measurement that is needed is totally unfeasible in practice today and probably for a very long time. As Leifer says (private communication):

> *Of course, For All Practical Purposes, it would be fine in any experiment we can do today for Wigner to assign a mixed state to the interior of the box. But interpretations of quantum mechanics have to deal with Impractical Purposes as well. They are supposed to tell us what is going on in any conceivable experiment allowed by the theory, so we cannot rely on FAPP. We are contemplating experiments where Wigner has the power to perform a coherent experiment on the box, including things like measuring whether it is in an entangled superposition state. Indeed, Universality implies that this can always, in principle, be done by some sufficiently powerful observer. We are assuming that Wigner predicts he would find the system in that entangled state with certainty, otherwise we would have to define what makes a system "classical", "an observer", or "a measurement device", which is precisely the kind of thing we are trying to avoid.*

From this point of view, the Wigner's friend scenario should not be interpreted as revealing an inconsistency within quantum mechanics, but rather as highlighting the limitations of classical assumptions about the nature of physical reality. More precisely, the tension arises from assuming that these different descriptions can be combined into a single observer-independent account. In particular, it challenges the idea that all facts can be embedded within a single, observer-independent framework.

We are now going to explore universality and its consequences in more detail.

# 4. Universality of QM

We will assume in the following that Universality applies and will examine the consequences of this assumption. This is not arbitrary: abandoning universality would require introducing an ill-defined boundary between quantum and classical systems, which is precisely what the measurement problem aims to avoid (as illustrated by Leifer's remark above).

The central question that we are going to address in the following is whether the universality of quantum mechanics can be maintained without abandoning the assumption of absolute events.

We adopt first-person universality, an excellent description of which is given by Adlam (2024)[7]:

[6] The first paper in which is introduced essentially the same idea is Zwirn (2016). However in ConSol there is no intrinsic state. It is mainly emphasized the fact that the components of the entangled state vector should not be interpreted as describing the actual physical state of the system in particular when the system is an observer.

[7] We don't consider here the notion of universality proposed by Ormrod and Barrett (2022) which is not relevant for what we want to show.

*The first [conception] is 'first-person' universality: to say that unitary quantum mechanics is 'universal' in this sense is to say that it is always a correct description of all the observations made by any individual observer, but it does not describe the relation between observations made by different observers. Here we should allow that the observations made by an individual observer, Alice, may include reports made to Alice by other observers about the outcomes of their own experiments, or records that she consults about observations made by other observers, which may be modelled quantum-mechanically as interactions between Alice and other observers/ records, or as measurements by Alice on other observers/records. Alice has no direct access to observations made by others, so her only information about them comes from the results of her interactions with other observers/records, and thus first-person universality requires only that Alice's own measurement outcomes are correlated with the results of Alice's measurements on observers/records in the way that quantum mechanics dictates -* ***first-person universality does not require that Alice's outcomes are correlated with the actual observations made by other observers in any particular way.*** *Therefore first-person universality places no constraints on the relations between observations made by two distinct observers, except in scenarios where there is some subsequent interaction or measurement in which the observers are able to share information about their outcomes*[8].

This means that the results are to be seen as linked to the observer who made the measurement. As Adlam says:

*Ultimately the only thing anyone can ever directly verify is the fact that QM makes the right predictions for a single observer.*

In contrast, the third-person point of view makes the additional assumption that the result obtained is an objective fact which can therefore be shared with other observers who will also be able to recognise it as a fact for themselves. This is the most standard intuitive point of view and it is adopted by realist interpretations. But this is in fact an additional hypothesis which is by no means obligatory and is not based on any serious reason other than our standard classical intuition.

First-person universality assumes the universal validity of unitary evolution, but restricts its application to the description constructed from a given observational perspective. It therefore gives a description of what is observed from the point of view of a particular observer who is outside the scope of this unitary evolution. It is by keeping the observer in question outside this unitary evolution that the measurement problem can be solved. The unitary formalism applies to all systems except the observer. Once an observer has obtained a measurement outcome for a system, they may invoke the reduction postulate (or the hanging-on mechanism in Convivial Solipsism) to adopt a new description of the system[9].

As we have seen, the universality of QM is sufficient in the simple context of the initial Wigner's friend experiment to derive what Adlam calls a Type II disaccord. This is what happens if Wigner, outside of the lab, adopts the state $\psi = \sum c_i |O_i\rangle|\varphi_i\rangle$ for his friend and the system. This state includes terms such as $|O_i\rangle|\varphi_i\rangle$ with different *i* while the friend inside the lab, having seen only one result $i_0$, should be considered as being in the state $|O_{i_0}\rangle$. From Wigner's point of view, adopting $\psi$ seems to mean that his friend has seen no result while his friend has seen one. In fact, from Wigner's point of view, the meaning of the components $|O_i\rangle$ inside $\psi$ should correctly be interpreted as meaning that if he asks his friend which result they got, he will hear the friend answering *i* with probability $|c_i|^2$. It is not that the friend is actually in the physical state of having seen the result *i*. So $\psi$ must not be interpreted as saying that the friend is in a superposition of having seen different results. Of course, it is counterintuitive to think that Wigner can hear the friend saying that they got the result *i* while the friend actually saw no result or a different result. This is because we have a tendency to merge the points of view of Wigner and the friend as if they were the same (as is the case in a usual third-person point of view).

The distinction between intrinsic state (supposed to refer to the actual physical state of the system) and interaction state (supposed to encode predictions about the results of measurements made on the system by a given observer in a certain context) provides a way to dissolve the apparent contradiction.

The universality of QM means that the unitary evolution applies to the state vector that is only supposed to enable us to calculate the probabilities of obtaining a given result if we make the measurement concerned. We must be even more vigilant in this respect when the state vector concerns a system which is itself an observer. The interaction state attributed to it must not be considered as describing its perceptions.

At this stage we can notice that the contradiction arising from Wigner's friend experiment, which essentially can be summarized in a type II disaccord, can be locally avoided within an epistemic framework, but only at the cost of abandoning universality or leaving the ontological status of states

[8] The last sentence of this description: "except in scenarios where there is some subsequent interaction or measurement in which the observers are able to share information about their outcomes" is in contradiction with the first part of the description and I do not accept it.

[9] Of course, as we mentioned before, this is not enough to solve the measurement problem unless what a measurement is can be specified (as is the case in Convivial Solipsism for example).

underdetermined[10].

But is this enough, as Adlam seems to think, to avoid concluding from this simple experiment that universality entails the non-absoluteness of events? In the case where events are considered absolute, if the friend has obtained a measurement result on a system that is considered as an absolute fact, we should be led to attribute to the system an intrinsic state which is an eigenvector of the measured observable. And it seems natural then to consider that the interaction state of this system which will give the best results for any observer is the same as the intrinsic state. In this situation, the interaction state and the intrinsic state should merge in contradiction with what universality implies. We could also say that the friend's result (if it is absolute) is for Wigner a kind of hidden variable (which is supposed not to exist). This strongly suggests that the assumption of absolute events is incompatible with universality. But we leave that aside because we are going to see in the following that the extended Wigner's friend experiments lead to difficulties that are much harder to eliminate.

# 5. Extended Wigner's friend experiment

## 5.1 The theorem

Many extended Wigner's friend experiments have been proposed recently. Although they are not all identical, they often share a similar experimental setup obtained by doubling the initial setup. There are two laboratories with a friend in each one and a "super observer" playing the role of Wigner outside each one. We will focus in what follows on the experiment and the no-go theorem proposed by Bong et al. (2020) which improves on a first, not entirely satisfactory, attempt made by Brukner (2017, 2018). We will use the simplified presentation given by Schmid et al. (2023) of this theorem. This experiment contains all that is necessary for our analysis of the assumption of universality[11]. The setup is the following:

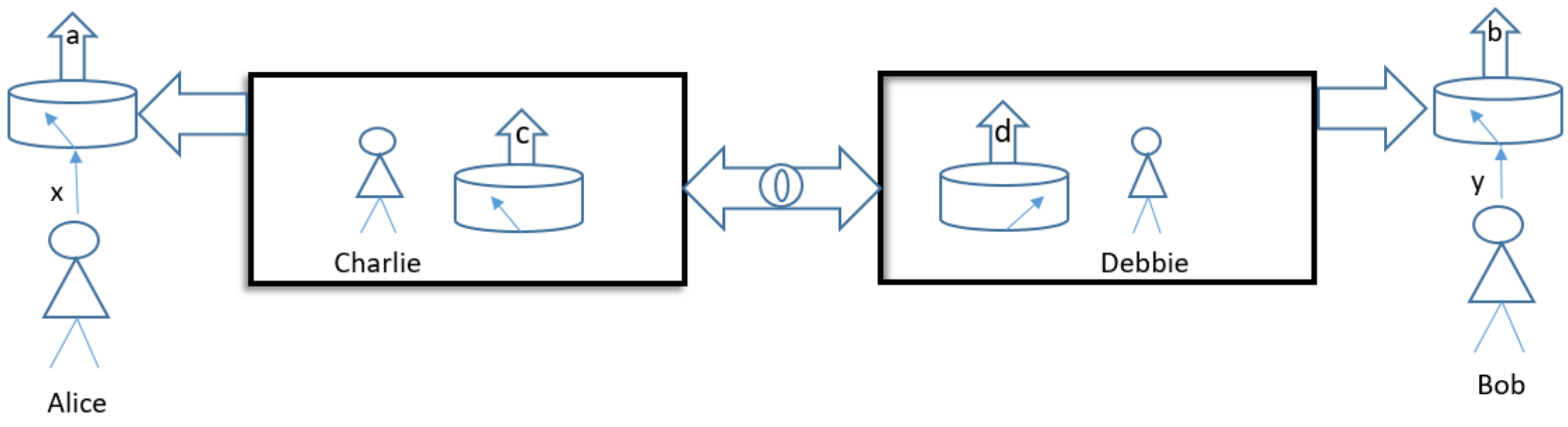

**Figure 1**: Charlie and Debbie inside their laboratories measure a pair of entangled particles and respectively obtain the results denoted c and d. Super observers Alice and Bob, outside the laboratories, perform spacelike separated measurements on the laboratories with settings x and y, respectively, and obtain outcomes a and b.

The following assumptions are made:

- Universality (U)
- Absoluteness of Observed Events (AOE)
- Local Agency for Observed Events (LAOE)
- A Super Observer can undo a measurement made by another observer (SO)

We described above what Universality is. AOE means that an observed event is a real objective event that is not relative to any observer or any situation. An observed event is then an event for all observers, even those who have not observed it. LAOE requires any freely chosen measurement setting to be uncorrelated with any set of relevant observed events outside that setting's future light cone. The requirement concerns joint distributions, not merely correlations with individual events considered separately. Roughly, the result an observer gets during a measurement cannot depend on the setting that another observer will choose in its future light cone.

The protocol is the following. Charlie and Debbie share a pair of particles entangled in the Hardy state: $\psi = \frac{1}{\sqrt{3}}\{|00\rangle + |01\rangle + |10\rangle\}$. They each measure their particle in the basis $\{|0\rangle, |1\rangle\}$. They obtain the results c and d. Alice can choose between asking Charlie which result he obtained (choice x=0) or undoing Charlie's measurement and measuring the particle in the basis $\{|+\rangle, |-\rangle\}$ (choice x=1). She obtains the result a. Bob can choose between asking Debbie which result she obtained (choice y=0) or undoing Debbie's measurement and measuring the particle in the basis $\{|+\rangle, |-\rangle\}$ (choice y=1). He obtains the result b.

It is easy to see that under AOE we have:

$$p(c=1, d=1|x=0, y=0) = 0 \quad (1)$$
$$p(c=0, b=-|x=0, y=1) = 0 \quad (2)$$
$$p(a=-, d=0|x=1, y=0) = 0 \quad (3)$$
$$p(a=-, b=-|x=1, y=1) = 1/12 \neq 0 \quad (4)$$

[10] Once more it is useful to insist on the fact that the measurement problem is not solved, however, since one still does not know which process leads to a determinate value.

[11] We refer the reader to the survey (Schmid et al., 2023) for a presentation and a comparison with the other experiments.

From these equations we can deduce under the LAOE assumption the following new equations:

$$p(c=1, d=1|x=1, y=1) = 0 \quad (5)$$
$$p(c=0, b=-|x=1, y=1) = 0 \quad (6)$$
$$p(a=-, d=0|x=1, y=1) = 0 \quad (7)$$
$$p(a=-, b=-|x=1, y=1) = 1/12 \neq 0 \quad (8)$$

The reason for switching from (1) to (5) is that since c and d are not in the future light cone of x or y, the probability distribution of c and d is not dependent on the choice of x or y. For switching from (2) to (6), the reason is that c and b are not in the future light cone of x so the probability distribution of c and b is not dependent on the choice of x. Similarly for switching from (3) to (7), a and d are not in the future light cone of y so the probability distribution of a and d is not dependent on the choice of y.

For a run with x=1 and y=1, the four outcomes a, b, c, d have been observed by at least one observer. The crucial point is that the AOE assumption implies that in this case a, b, c, d can be considered as absolute events forming a joint outcome. It is then possible to reason from equations (5) to (8) and to chain them by transitivity. From (5), we see that d=1 => c=0. From (6), c=0 => b=+. From (7), a=- => d=1.

By transitivity, a=- => d=1 => c=0 => b=+. Hence a=- => b=+ in contradiction with (8).

The contradiction relies on treating the four outcomes as jointly defined and combining them through transitivity. The four assumptions (U), (AOE), (LAOE), (SO) are not compatible and it is necessary to abandon at least one of them.

## 5.2 Which assumption should be discarded?

We see that the four assumptions are not mutually compatible so at least one of U, AOE, LAOE and SO must therefore be abandoned. As we said before, we restrict ourselves to analysing interpretations inside a non-modified formalism so we do not speak of spontaneous collapse interpretations or of dBB theory. We also do not discuss the Everett interpretation since we assume that there is only one outcome obtained by the observer during each measurement.

Universality is rejected by interpretations that postulate that the unitary evolution does not apply to the measurement apparatus or to the observer: Copenhagen interpretation (Faye, 2019), CSM (Auffèves & Grangier, 2016; Van Den Bossche & Grangier, 2023) and so on. But none of these interpretations succeeds in giving a good criterion to decide which systems escape universality[12]. This is the difficulty highlighted by Leifer in the passage quoted in Section 3.

This is the reason why the interpretations that he describes as "Copenhagenish" all assume Universality as a basic postulate. In the following, we will situate ourselves inside the framework of Copenhagenish interpretations. If we keep U, it seems difficult to refuse SO[13]. Of course, the fact that a Super Observer can undo a measurement made by a friend is often considered as unacceptable. It is true that it is something very far from our current possibilities and it will probably stay beyond any human capabilities in the future. But if we accept Universality then, in principle, this is something that is allowed by physics. So we have to take its consequences into consideration. We are naturally led to focus on the choice between discarding AOE or LAOE.

Nevertheless it can be noticed that the conjunction of AOE + Universality leads to a tension because it is not clear how it is possible that when an observer carries out a measurement they obtain a single and absolute result if the observation is a unitary process[14]. This seems possible only in perspectival interpretations where outcomes are relative. In this case, as we will see, the measurement is described as a unitary process by an external observer but is not unitary for the observer who performs it. We will come back to this point in sections 6 and 7. Actually perspectival interpretations reject AOE.

There is also a problem between AOE and SO because if a super observer is able to undo a measurement, the result that had been obtained is erased, and therefore it makes no sense to consider that it is an absolute event and that we can use it as if it still existed to compare it with results obtained after undoing the measurement. This seems similar to measuring the spin along Ox, then undoing it and measuring the spin along Oz, and then using the joint probability of two results corresponding to incompatible observables. We will come back to this in section 6.

These tensions do not invalidate the theorem, but show that its interpretation depends critically on how its assumptions are understood. We will return to these tensions later, after analysing the structure of the assumptions in more detail.

Adlam does not want to accept the metaphysical implication of the non-absoluteness of observed events, so she keeps AOE and proposes a retrocausal mechanism that violates LAOE. We will see that one of the reasons she refuses the non-absoluteness of observed events is that she thinks that it would endanger our scientific methodology. We will argue below that this is not the case.

Now many interpretations keep U and LAOE and choose to abandon AOE. What does that mean? It is easy to understand the meaning of AOE. It corresponds to the usual way we consider events. It is less easy to understand what the fact that events are not absolute could mean. As Adlam rightly notices, in the papers about Extended Wigner's friend, the non-absoluteness is simply defined as the negation of AOE. But it is important to understand what non-absoluteness looks like[15]. This is exactly the question that we are going to deal with in the following.

[12] Grangier (Auffèves & Grangier, 2016) is working with the hope of providing such an explanation through the use of type III von Neumann algebras but this is currently only work in progress and nothing guarantees that it will work.

[13] We will see that this is nevertheless what is done in ConSol but the reason comes from the way the measurement is described.

[14] This is not possible in standard quantum mechanics since we start from a superposed state vector. Things are different in dBB theory where this problem does not happen.

### 5.3 Type I and Type III disaccords

In this section we will analyse what happens when AOE is not respected. Events are non-absolute in relational or perspectival interpretations which state that the result of a measurement is relative to the observer who performed the measurement. So there is no fact of the matter about which outcome actually occurred. This is the case in Relational Quantum Mechanics, in QBism, in the modal interpretation, in Convivial Solipsism, in Brukner's account and in many other interpretations sharing this view that Adlam calls "the metaphysically radical non-absoluteness". Let us emphasize the fact that in this kind of interpretation an observer has no direct access to the result of another one. Recall Adlam's description of the universality in the first person that we quoted in section 4.

There are many ways to understand what it means for a result to be relative to an observer. The differences appear mainly when we wonder what happens when two observers compare their results and they depend on the procedure they will use for the comparison. They depend also on the metaphysical framework that is adopted. By metaphysical framework I mean not only the ontological commitment but also the way we allow ourselves to speak of the interpretation itself. It is similar to what logicians call the meta-level of a formal axiomatic theory. A formal axiomatic theory allows us to express sentences that are correct inside the theory and can be true or false. A proposition P can be proved and it becomes a theorem but saying that the proposition P is true is not possible inside the theory itself. The sentence "the proposition P of the theory is true" is not a sentence of the theory but a sentence of the metatheory.

Explicating an interpretation of a formalism means speaking at the meta-level. In a standard realist world, one can use everyday language to talk about what the theory says and most often, there will be no ambiguity. But we will see that this is no longer the case in the strange world of perspectival theories and that we will have to be more precise because the asserted sentences can no longer be considered as generally and neutrally spoken but must be associated with a designated speaker, what we call a perspective, and this point changes everything. A meta-level description is a theoretical construct used to compare or relate different perspectives. It does not correspond to the viewpoint of any physical observer and does not provide access to a global state of affairs. Its role is purely formal: as we will see, it allows one to express relations between perspectival descriptions (for instance, to formulate assumptions such as Tracking), without implying that these relations are accessible within any single perspective.

While these issues are conceptually important, our central concern will be the role of the Tracking assumption (defined below) in the no-go argument.

We have analysed above the type II disaccord and found a way to make it harmless. Following Adlam once again (we have just changed the forenames she uses), we now turn to two other more severe types of disaccord.

- Type I disaccord

Charlie performs a measurement and obtains a result C1. Alice performs a measurement and obtains a result A1. Alice infers from the result A1, using quantum mechanics, the result of Charlie, and it does not correspond to C1.

- Type III disaccord

Charlie performs a measurement and obtains a result C1. Alice performs a measurement on Charlie by asking him which result he obtained and gets a result A1. Alice then infers from A1 the result of Charlie by intuitively interpreting Charlie's response, that is, by inferring that the result obtained by Charlie corresponds to the answer, namely A1, and it does not correspond to C1.

### 5.4 On the possibility of comparing results

A first question is: in both cases, can this disaccord be proved empirically? Who is able to compare A1 and C1? Who is uttering the sentence "A1 does not correspond to C1"? Neither Alice nor Charlie can do that because neither of them has any access to the actual result of the other.

The very concept of relativity of results needs the possibility of comparing different results. To say that the result of a measurement is relative to the observer who makes it means either that no measurement has occurred for another observer or that the same measurement could provide a different result. It is therefore only possible to express this concept of relativity if we accept the possibility of comparing the values obtained by different observers. It then becomes essential to specify the meaning given to sentences as simple in appearance as "the results of Alice and Bob are different". Who is the legitimate speaker of this sentence? Many papers deal with this subject in a superficial way, with the consequence that they state propositions containing implicit presuppositions which they neglect but which are contradictory to the conclusions they claim to draw.

In perspectival interpretations where an observer has no direct access to the results of another observer, uttering this sentence is possible only at the meta-level, understood not as a God's-eye viewpoint but as a theoretical language for comparing indexed perspectival claims. The meta-level is a theoretical construction enabling us to speak of the way the interpretation works. It is not the standpoint of any real observer. Allowing us to speak from this meta-level point of view is necessary; otherwise it is impossible to express what these disaccords mean. This is a generalisation of what Pienaar (2024) calls "the spectator"[16].

[15] Adlam says that, after all, it is possible that there is no coherent way in which events could fail to be absolute. Then the relevant theorem would have proven (if we do not want to let go of the other assumptions) that quantum mechanics is not universal. But we will argue that it is possible to understand what non-absoluteness means.

## 5.5 Iteration of relativization

Riedel (2024, in press) wonders whether the relativization can be iterated. That raises strange questions and makes a meta-level description even more necessary. Let us try to understand how this concept of iteration of relativization could be understood.

The notion of relativity of a measurement result to the observer who makes the measurement is relatively natural. Bob makes a measurement that produces a result r. Saying that the result r that Bob obtained is relative to Bob means simply that Alice could obtain a different result for the same measurement. The notion of second-order relativity, which, according to Riedel, concerns the fact that the sentence "the result r that Bob obtained is relative to Bob" is itself relative, is more difficult to understand. One way to understand it leads to the concept of relativization of scenarios. If the fact that "Alice is in a lab making a measurement" is relative to Wigner, then "the result that Alice obtained relative to Alice" is relative to Wigner because it is possible that from Alice's point of view she was not in the lab and did not make any measurement. What does that mean? For Wigner, Alice made a measurement that is relative to her because the result she got in this scenario (which is Wigner's scenario) is not a result that Wigner can know so Wigner could get a different result for the same measurement. Now for Alice, perhaps no measurement has been done because she was not inside the lab in her own scenario or she did a different measurement. So "the result Alice got relative to Alice" is itself relative.

If we consider three observers, things become even more complicated. To say that "the result that Alice got relative to herself" is relative could mean that for Wigner1, Alice got a result r in a scenario1 for Wigner1 where she did a measurement and that r is relative to Alice because Wigner1 cannot know it and could get a different result if he were doing the same measurement. Now for Wigner2, Alice got a result r' in a scenario2 for Wigner2 where she did a measurement and r' is relative to Alice because Wigner2 cannot know it and could get a different result if he were doing the same measurement. Now, r and r' (which are both results relative to Alice but in different scenarios) can be different. It could even be possible that the result r that Alice got in Wigner1's scenario be related to a z-spin measurement while the result r' that Alice got in Wigner2's scenario be related to an x-spin measurement, or that Alice made no measurement in Wigner2's scenario. That helps us understand what it means to say that "the result obtained by Alice relative to Alice" is itself relative. Moreover, it need not be the case that Alice obtained "a result relative to herself" relative to herself since for her, it is possible that she did not make any measurement. That is the way we can understand the fact that "the result that Alice got relative to herself" could be relative, what Riedel calls iteration at second order.

If we fully accept the idea that quantum mechanics is universal and that even the macroscopic world is governed by unitary transformations, then we must accept that the environment we perceive also arises from quantum measurements. In this case, the scenario in which we live (what we call our everyday reality) is just as relative as the result of a spin measurement.

Then we could even say that this seeming second-order relativization is not different from the first-order relativization since, after all, it is only a first-order relativization generalised to the macroscopic world. The key step is to recognize that perceiving our environment is not different from making a measurement of a quantum property. So we are led to consider that refusing this generalised relativization is not consistent if we fully accept the relativization of results coming from usual quantum measurements.

This generalisation, which, in light of the preceding considerations, I prefer to call "first-order relativization" rather than "second-order relativization", leads one to conclude that "the result r relative to Alice" is itself relative to the observer who states it.

Riedel says that the first versions of RQM must imply the iteration of relativization. According to him, more generally if we accept second-order relativization there is no reason to refuse to iterate again even though relativization at any order is something that becomes rapidly impossible to describe.

So do we need to continue iterating relativization indefinitely in every case? Not necessarily[17]. Let us adopt the following convention: For a given observer, the scenario in which they live and the measurement results they obtain are relative to themselves and both relativities are first-order relativities[18].

Actually the level of iteration that is needed depends on the situation considered. If we consider a hierarchy of nested observers, where at least one observer describes the viewpoint of another, who is in turn described by a third observer and so on, the number of levels of iteration that is needed is equal to the number of nested observers taken into account minus one. When only one observer is considered (say Alice alone) then there is no relativization. From Alice's point of view everything seems absolute. If we include another observer (say Bob who is speaking of Alice's results) then we have to consider that Alice's results are relative to her because Bob could obtain other results for the same measurements and moreover, the scenario for Bob can be different from Alice's scenario. Alice can perceive herself as being in a lab performing a z-spin measurement while Bob can perceive that she is performing an x-measurement or no measurement at all. So with two observers we need to relativize both the results and the scenarios. This is a first-order relativization. Now if we add Charlie who just plays a similar role to Alice (Charlie is not speaking of Alice's results, Alice is not speaking of Charlie's results and Bob is not speaking of Alice speaking of Charlie's results) then the first-order is still enough. However if Bob is going to mention what Alice heard about Charlie's results then we switch to the second-order relativization because we have three nested observers. I will not describe the situation with four observers but readers will be able to generalise the reasoning for themselves.

[16] Pienaar (2024) says "*in any given experimental run, we would like to assert that the measurement settings of all observers simultaneously take definite values, that is, we can imagine a list that tells us precisely what values of x, y, z, . . . are chosen in a given run. However, phrased in this way, it is not clear that any actual observer could verify that this assertion is true: it seems to appeal to a 'view from nowhere', which is counter to the spirit of perspectival interpretations. To avoid this problem, we shall explicitly include an ultimate observer whom we shall call 'the spectator', whose role is merely to collect and record data about all of the other observer's measurement settings. Specifically, we assume that in each run all observers are required to communicate (say, via classical channels) their choices of setting variables x, y, z, . . . to the spectator, who then maintains a record of their values, over all runs, until the end of the experiment.*" But this spectator cannot be any actual observer because assuming Universality means that there are no classical channels. So depending on whether the Tracking assumption (see below) is violated or respected, what the spectator collects will or will not be consistent with the real results.

[17] We reject the possibility (suggested by Riedel) of relativizing to the same reference multiple times which is difficult to justify and which does not make sense in ConSol.

[18] I emphasize that contrary to Riedel, I consider that relativization of "the result r is relative to Alice" is not second-order relativization but only taking into account the relativization as first-order relativization of scenarios.

We see that, in contrast with RQM where observers can be any physical systems and so the relativization seems to never end, if we consider observers as perceiving entities, we can set a limit on the number of levels of iteration which depends on the number of observers considered.

### 5.6 Are second-order facts absolute?

Let us call the fact that "the result r is relative to Alice" a second-order fact. Are second-order facts absolute, as the tame relativist accepts according to Riedel (in press)?

> *"The tame relativist accepts absolute facts about what is the case relative to this or that observer. If we were not allowed to 'juxtapose' these facts in any way, relative tracking could not even be formulated."*

As we have shown above, this is not the case when at least two nested observers are considered. But it is true that we have reasoned as if the Tracking assumption, introduced in Schmid et al. (2023), and analysed in detail below, were not respected.

There are two possible ways to analyse second-order facts depending on whether or not the Tracking assumption is accepted.

1) One way to make sense of this absoluteness of the second-order facts is to accept the Tracking assumption, which allows one to link what Alice actually perceived to what another observer hears her say she perceived. The Tracking assumption states that when Alice directly asks Charlie which result he obtained, what she hears is really what Charlie obtained. This can be mathematically written (if we use the same notation as before): $p(a|c, x=0) = \delta_{a,c}$. Of course the Tracking assumption automatically eliminates type III disaccords. The Tracking assumption is equivalent to what Adlam calls Cross-Perspective Communication. Under the Tracking assumption "a result r relative to an observer" is not relative since the observer can communicate their result to other observers who then share the result. So in this case, it is not clear how the result itself is still relative. It seems that results become absolute because they can be shared between all the observers. Actually it is not totally the case. If Charlie obtains a result relative to him, it is not a result for Alice until she asks Charlie which result he obtained. That means that before asking Charlie, Alice should use an entangled state vector for Charlie and the system while after having heard Charlie's answer, she will use a state vector limited to the result she heard. So under the Tracking assumption, results are relative to the observer who made the measurement and become results for another observer only after the latter asks the former for the result they obtained. Let us emphasize that this is a partial relativism limited to the fact that a result for an observer may not be a result for another observer (for the latter there is no result)[19]. Remember that a total relativism allows two observers to obtain different results, which is stronger and not allowed here.

It is not clear if the Tracking assumption also eliminates type I disaccords. It does not seem that this is necessary. In the case of a type I disaccord with Tracking respected, a strange situation could happen: Charlie first performs a measurement and obtains a result C1. Alice then performs a measurement and obtains a result A1. Alice uses quantum mechanics and infers from her result that Charlie obtained a certain result. Now she interrogates Charlie who tells her that his result is different. It seems that this situation is even worse than what arises in type III disaccords. Indeed in the case of a type III disaccord, Alice hears Charlie saying that he obtained A1 while in reality he obtained C1 ≠ A1. But Alice has no way to witness the contradiction because she has no access to Charlie's actual result. The type III disaccords can only be expressed from the meta-observer point of view. They are not directly in conflict with the experience of any observer. On the contrary, in the case of a type I disaccord with the Tracking assumption respected, Alice knows the result she rightly inferred and she also knows the real result Charlie obtained and although they should be identical, they are different. This is a much stronger conflict! A useful analogy is the following: it is like hearing a friend report an event that you have no independent access to, versus hearing a report that directly contradicts something you have yourself observed. The latter case represents a much stronger form of inconsistency. So assuming the Tracking assumption is not enough to restore genuine intersubjectivity unless type I disaccords are also eliminated, which is of course the case for absolute events but not necessarily if events are relative.

---

[19] It seems that it is Cavalcanti's position.

Moreover, as Riedel shows, respecting the Tracking assumption exposes the interpretation to revenge theorems (see section 5.8). So we see that accepting the Tracking assumption does not make second-order facts absolute in a perfect sense since even if "the result relative to an observer" is not relative to another observer, it may not exist for another observer before they interrogate the first one. Second-order facts are absolute only at the meta-level.

2) If the Tracking assumption is rejected we saw that the level of iteration depends on the number of nested observers which are taken into account. So second-order facts cannot be absolute as soon as at least two nested observers are involved. But there is nevertheless a way to use second-order facts to state the Tracking assumption which compares Alice's result r and Charlie's result r'. Let us remember that the Tracking assumption states that when Alice directly asks Charlie which result he obtained, what she hears is really what Charlie obtained. We have seen that in the case where only two nested observers are involved, the necessary number of iterations of relativization is only one. In this case, we just have to consider that the results that are mentioned in the Tracking assumption are all results in Alice's scenario. Then it is possible to state the Tracking assumption in a meaningful way even if second-order facts are not absolute by specifying that the results mentioned in the Tracking assumption are those obtained in the scenario of the observer who asks the question. Of course, only the meta-observer is able to notice this.

The meta-level makes it possible to mention the results obtained by two different observers and to compare them. Some authors[20] reject the meta-level description, but this prevents a meaningful formulation of type I and III disagreements. Moreover not being able to state that the results from Alice and Charlie are different does not imply that they are the same. However these authors sometimes state that, since it is not possible to even state that the results are different, this allows one to consider that when Alice hears Charlie's answer what she hears corresponds to what Charlie really obtained. They then think they have regained a kind of intersubjectivity. But this position is not acceptable.

## 5.7 The Tracking assumption: discussing the different positions

The Tracking assumption is a central and very subtle point on which authors' positions differ. This issue has been discussed, without this terminology, in previous work on Convivial Solipsism, starting with Zwirn (2016). The difficulty comes from the fact that at first sight, it seems totally obvious that it should be respected, because it is so deeply ingrained in our standard habits[21]. Denying the Tracking assumption is considered by Schmid et al. (2023) as quite a radical position. Bong et al. (2020) consider that the Tracking assumption is a methodological necessity and Adlam (2024) wants a Cross-Perspective Link (which is another name for the same idea) to recover intersubjectivity. So it is interesting to see if the Tracking assumption is needed for the proof of the theorem or if it is possible to prove it without this assumption.

In what follows, we will interpret the Tracking assumption as the requirement that the outcome that an observer hears reported by another observer coincides with the outcome actually obtained by the latter observer.

The relationship between AOE and Tracking depends on the precise definition of AOE. If absoluteness includes faithful transmission between observers, Tracking is included by definition. If AOE states only that observed outcomes have observer-independent values, faithful transmission requires an additional assumption.

In the stronger reading, AOE includes both definite outcomes and faithful transmission between observers. Tracking then follows from the definition being adopted, rather than from observer-independence alone.

Schmid et al. (2023) distinguish these commitments. Besides the Parallel Lives example (Brassard & Raymond-Robichaud, 2020), which illustrates Tracking without AOE, they envisage absolute outcomes followed by reports that do not reproduce those outcomes. Their claim of independence therefore addresses both directions.

One may question the physical or epistemological significance of absolute outcomes that are not faithfully transmitted. However, this objection must be distinguished from the claim that the weaker definition of AOE logically entails Tracking.

Okon (2025) says that AOE has two independent parts: 1) the claim that there is a joint probability distribution for all four results and 2) the Tracking assumption. He says that only the first part is required to enforce absoluteness of observed events and that the Tracking assumption is by no means obvious:

> *The key issue to have in mind is that, even when Wigner and his friend measure "in the same basis", it is never the case that what they are comparing are the result of the exact same measurement (I use the quotations to emphasize the fact that friends and superobservers perform measurements on different systems so, strictly speaking, the bases cannot be the same). As a result, a demand for absoluteness of observed events does not necessarily imply a demand for these two measurements to yield the exact same result.*

This claim requires clarification. Actually Wigner can measure "in the same basis" as the friend in three different

[20] Cavalcanti for example refuses to consider second-order facts: "It is just the rules of quantum mechanics that if Wigner measures the system and sees spin up then asks the friend what he saw, he will say "spin up". There's still no risk that the friend "really" saw spin down, because, remember, there's no second-order fact about the friend "really" saw." (Private communication). I disagree with this position.

[21] Cavalcanti considers that speaking with somebody when the Tracking assumption is not respected is equivalent to speaking to a zombie. (Private communication). For me it is a misunderstanding of what it means to deny the Tracking assumption in a perspectival interpretation.

ways. Schmid et al. comment on Okon's quotation as follows:

> *Specifically, when x=1, if Alice obtains the outcome for the computational basis measurement on the system R not by asking Charlie, but by reversing Charlie's measurement and then herself measures the system R in the computational basis, then there is no guarantee that she will get the same outcome as Charlie. Even though the effective dynamics of Charlie's measurement on R followed by its reversing is represented as an identity channel on R in quantum theory, in interpretations that violate transformation noncontextuality (such as Bohmian mechanics) or that have inherent stochasticity, it is entirely possible that Alice's outcome will not be the same as Charlie's outcome. Still, as long as Alice actually implements her measurement by asking Charlie what he observed (as in the original Local Friendliness proposal), then Eq. (48) as an instance of Tracking is hard to deny.*

Schmid et al. acknowledge the fact that if Wigner undoes the friend's measurement and measures the system in the computational basis, it is not necessary for him to get the same result as his friend. That is the easy part because first, there is the possibility that the system has been modified in some way by this reversal and second, if the measurement is stochastic there is no reason for the second measurement to provide the same result as the first. Now it is also possible that Wigner does the measurement in the same basis as his friend without undoing the friend's measurement (assuming that the first measurement was non-destructive and that no perturbation happened in the meantime). They just say that in this latter case the Tracking assumption is hard to deny. On the contrary, for reasons explained below, my position is that there is no necessity for Wigner to get the same result and no reason for the Tracking assumption to be respected in either case.

Like Adlam, Okon states that AOE is equivalent to the conjunction of the Tracking assumption and another hypothesis but he replaces Adlam's AOE1 (which states that there is a well-defined value for the outcome observed by each observer) by the existence of the joint probability of the four results observed by Alice, Bob, Charlie and Debbie. It is not obvious that these two assumptions are equivalent.

A first difficulty is to sort out how these four hypotheses are interconnected:

- Events are absolute and not relative to any observer.
- There is a well-defined value for the outcome observed by each observer.
- The joint probability of the four results observed by Alice, Bob, Charlie and Debbie exists.
- The Tracking assumption.

A second difficulty is to determine which of those are necessary to prove the no-go theorem.

Adlam argues that the Tracking assumption is necessary to obtain the result in Bong et al. (2020). By the way, the Tracking assumption is explicitly assumed in Bong et al. (2020). Indeed Adlam uses it to establish that the four outcomes exist in a single reality, thereby allowing the switch from Eqs. (44)-(47) to Eqs. (32)-(35)[22]. This is one way to get this result. Indeed, starting from the quantum predictions:

$$p(a=1, b=1|x=0, y=0) = 0 \qquad (9)$$
$$p(a=0, b=-|x=0, y=1) = 0 \qquad (10)$$
$$p(a=-, b=0|x=1, y=0) = 0 \qquad (11)$$
$$p(a=-, b=-|x=1, y=1) \neq 0 \qquad (12)$$

and using the Tracking assumptions $p(a|c, x=0) = \delta_{a,c}$ and $p(b|d, y=0) = \delta_{b,d}$ we obtain Eqs. (1) – (4). Applying LAOE then yields Eqs. (5) – (8) from which the contradiction follows under the assumption that the four outcomes admit a joint probability distribution. Then she argues that to escape the contradiction we arrive at, it is enough to relax AOE2 while keeping AOE1. She concludes that this is a proof that this kind of experiment leads to type III disaccord.

But Schmid et al. state that the Tracking assumption is not necessary for the proof. According to them, equations (1), (2), (3) and (4) come directly from quantum mechanical predictions and concern results of measurements directly made by one of the observers. The contradiction follows from equations (5), (6), (7), and (8) that can be obtained from equations (1), (2), (3) and (4) by using only LAOE without any need for the Tracking assumption.

If we follow Okon who states that AOE is equivalent to the conjunction of the Tracking assumption and the hypothesis "The joint probability of the four results observed by Alice, Bob, Charlie and Debbie exists", then it is true, as Schmid et al. say, that the theorem can be proved without the Tracking assumption since it is possible to start directly from Eqs. (1)-(4).

So, according to Schmid et al., Adlam's proof that 1) abandoning AOE2 is enough to escape the contradiction and 2) the experiment leads to type III disaccord is not justified.

> *But in any case, our presentation of the Local Friendliness argument makes it clear that the specific instance of the Tracking assumption in Eq. (48) is not necessary, so one cannot avoid the no-go argument by challenging it. To see this, note that this condition was only required to get from Eqs. (44)-(46) to Eqs. (32)-(34), but these latter equations are themselves directly observable predictions of quantum theory. So one can begin the argument from this step, after which*

[22] Eqs. (44)-(47) and (32)-(35) in Schmid et al. (2023) correspond respectively to Eqs. (9)-(12) and (1)-(4) in this paper.

*the argument makes no reference to Eq. (48)*[23].

But these positions must be compared using the same definition of AOE. Rejecting Tracking rejects a stronger assumption that includes Tracking, but does not by itself reject the weaker assumption that the observed outcomes admit a joint probability distribution. A proof that uses only this weaker assumption is therefore not invalidated merely by rejecting Tracking.

The absence of an observer who can access all four outcomes raises a question about the empirical interpretation of their joint distribution. It does not, by itself, establish that such a distribution is mathematically undefined. To challenge the derivation, one must identify the particular probabilistic premise or inference that the interpretation rejects.

We will not reach a definite conclusion about which author is right in this debate since ConSol rejects the treatment of outcomes from different perspectives as observer-independent components of a single joint outcome. Its response to the theorem therefore rests on rejecting AOE in this sense. Moreover the rejection of Tracking is a further feature of its account of communication.

Hence, ConSol, rejecting both AOE and the Tracking assumption, escapes the contradiction.

### 5.8 Tame and feral interpretations

We presented Riedel's ideas about iterated relativism (Riedel, 2024, in press). He also introduces an interesting distinction between the different relativist interpretations. This distinction can be related to the discussion in the previous section.

Riedel distinguishes between two kinds of relativist interpretations: the tame and the feral.

| | Tracking assumption respected | Tracking assumption violated |
|---|---|---|
| Second-order Absoluteness | tame | feral |
| Higher-order relativity | feral | feral |

Fig 2. Tame and Feral quantum relativism

He shows that for interpretations which have second-order absoluteness and respect the Tracking assumption, it is not enough to reject AOE to escape the no-go theorem. Even without AOE it is possible to write a slightly modified form of the theorem (that he calls a revenge theorem) which applies. So the only possibility seems to be to become feral either by accepting higher-order relativity or giving up the Tracking assumption and the Cross-Perspective Link or both. This is precisely the strategy adopted by Convivial Solipsism, which rejects both AOE and the Tracking assumption, thereby avoiding both the original no-go theorem and its revenge variants.

## 6. A series of questions: must we accept the theorem?

We have already mentioned some reservations about the non-trivial conjunction of some of the hypotheses that are used in these theorems. In order to follow the reasoning through to the end, we have proceeded as if this were not problematic but we now clarify these problems that can lead some physicists to reject the whole argument.

Of course, one hypothesis is not very credible: the possibility of a super observer undoing a measurement made by another observer (SO) is not something that is easy to accept. It is true that assuming Universality seems to lead to SO at least in principle. But the impossibility of doing it in practice (not only currently but probably also in the future) makes it a bit unclear for many physicists. The same problem arises with the measurement of the whole system including the friend, the system and the lab, which is totally unfeasible in practice. The authors of the theorem propose a prospective experiment (Wiseman et al., 2023) that, even though out of reach currently, could be done in the future. I will not analyse this experiment further here. ConSol distinguishes a super-observer's control of physical interactions from any purported reversal of another observer's first-person experience.

Another reservation concerns the conjunction of Universality and the fact that a measurement provides a unique result. If a measurement is a physical process described by a unitary interaction, it is difficult to understand how it can provide a unique result, at least in standard quantum mechanics where superposition is preserved by unitary transformation[24]. This problem is often bypassed by authors who adopt both assumptions without any further questions. Actually the only way to make sense of it is to separate the measurement on a system S giving a unique result, which must be done by one observer O for whom the measurement is not a unitary process and the description by another observer O' of the measurement of S by O. The latter description can then be unitary. So Universality concerns the description of a measurement when the observer and the system (the apparatus, the environment etc.) are considered as a whole from the point of view of an external observer. The unitary description applies to the external modelling of the measurement process, while the appearance of a definite outcome is perspectival and not described by a physical transformation, so it is not unitary. This clarification is essential. It is totally compatible with the description of a measurement given by Convivial Solipsism as described in section 7.2. The distinction between the physical interaction, which is unitary, and the fact of obtaining a result, which is not[25], clarifies the issue.

[23] Eqs. (44)-(46) and (32)-(34) in Schmid et al. (2023) correspond respectively to Eqs. (9)-(11) and (1)-(3) in this paper. Their Eq. (48) expresses the relevant Tracking conditions.

[24] As we said in section 5.2, things are different in dBB theory where this problem does not happen.

Another difficulty comes from the conjunction of AOE and SO. Absoluteness of observed events means that a result that is obtained by an observer from a measurement is absolute and so, it is not only a result for the observer who obtained it, but also a result for any other observer (with the same actual value) and it will stay as a result in the future. If Alice obtained (in an absolute sense) a result r at a time $t_0$, the result r is a result for everybody and the fact that Alice obtained this result r at the time $t_0$ will always be true. But SO implies that it is possible to undo this result at time $t_1 > t_0$ and to erase everything about this result including Alice's memory. The issue is not merely the deletion of a record: the inverse unitary restores the pre-measurement physical state, in which no outcome had yet been recorded. I take this restoration to undermine the treatment of the undone outcome as an absolute fact that remains available for an observer-independent joint assignment. Maintaining that the outcome nevertheless survives as a historical fact is precisely the additional ontological commitment I question. In this case, it is difficult to see how the result r can be considered absolute. The only possibility would be to consider a new concept of absoluteness that is linked to a period of time: the result r is absolute between the time $t_0$ and the time $t_1$.

But in this case, it seems problematic to use a joint probability with a result r that is absolute during a certain period (before it is undone) and a result r' which is obtained outside of this period. However that is precisely what is done in the no-go theorem when we use joint probabilities between the result that Charlie obtained and the result that Alice obtained after having undone Charlie's result (in the case x=1). If we recall the proof of the theorem, it uses the four joint probabilities (5) – (8) given in section 5.1. They involve both Charlie's earlier outcome c and Alice's later outcome a, obtained after reversing Charlie's measurement interaction. My objection concerns treating the earlier outcome, whatever its value, as an absolute fact available for a joint assignment after the measurement has been undone[26]. As we will see, the interpretation of these joint probabilities assumes the coexistence of outcomes that ConSol rejects.

Nevertheless, rather than questioning the validity of the theorem, we take it seriously as a powerful constraint on any interpretation satisfying its assumptions. It is therefore necessary to identify which of these assumptions must be abandoned. The principal point is that it allows us to explore interesting questions which, I contend, are all answered by Convivial Solipsism that we are going to present now. In the next section, we show that ConSol provides a coherent framework in which the contradiction is avoided by rejecting the Absoluteness of Observed Events.

# 7. How feral is Convivial Solipsism (ConSol)?

Convivial Solipsism and the way it solves many puzzling problems in quantum mechanics such as the measurement problem, the EPR correlations and the associated non-locality, the delayed choice experiments (etc.) have been presented in a series of papers (Zwirn, 2015, 2016, 2020, 2021, 2022, 2023, 2024, 2025, 2026)[27]. I analyse in this section how Convivial Solipsism behaves with respect to these no-go theorems. I start by explaining the main concepts on which ConSol rests. The whole interpretation comes from a detailed analysis of what a measurement is.

## 7.1 Core framework and definitions

We introduce three key notions: *perspective*, *observer*, and *meta-level description*.

A perspective P is defined as a temporally ordered sequence of recorded outcomes, corresponding to the successive perceptual experiences of a single observer. Each perspective provides a coherent account of measurement results constituting the phenomenal reality of the observer, and all statements about outcomes are meaningful only relative to such a perspective.

An observer is a stable structure with the ability to engage in interactions that produce further outcomes attached to a perspective, characterized by memory and consistency.

A meta-level description is a theoretical construct used to compare or relate different perspectives. It does not correspond to the viewpoint of any physical observer and does not provide access to a global state of affairs. Its role is purely formal: it allows one to express relations between perspectival descriptions (for instance, to formulate assumptions such as Tracking), without implying that these relations are accessible within any single perspective.

This distinction is crucial. Statements comparing outcomes as experienced within different observers' own perspectives require a meta-level formulation. A statement within one perspective may nevertheless concern several observers or several outcomes, provided that these are described as they appear within that perspective.

[25] In contradiction with what Schmid et al. write "the outcome obtained when an observer carries out a measurement process is single and absolute **(even if the observation is part of a unitary process)"**, we say that observation is never part of a unitary process!

[26] Recall the difficulty we noticed in section 5.2 about the fact of measuring the spin along Ox, then undoing it and measuring the spin along Oz, and then using the joint probability of the two results corresponding to incompatible observables.

[27] Convivial Solipsism was developed in 2000 as a detailed extension of a remark made by d'Espagnat in his first book (d'Espagnat, 1971) criticizing the Everett interpretation. Much later, the attention was drawn by an anonymous referee of Foundations of Physics and by Jeffrey Barrett (1999) to similarities with the single-mind theory of Albert and Loewer (1988). Nevertheless there are many differences between the two interpretations.

## 7.2 What is a measurement?

In ConSol, a measurement is not identified with the unitary interaction between a system and an apparatus, but with the perception of a definite outcome by an observer. Measuring a property of a system requires interacting with the system using a macroscopic device in such a way that the possible values of the property of the system to be measured correlate with a quantity of the device that is directly perceptible to an observer. For example, a spin measurement of a particle will be coupled with the position of a needle on a dial or the impact of the particle on a screen or the click of a detector. The position (via a light emission or a black spot printed at the point of impact) and the click (via the noise produced) are macroscopic quantities that are directly perceptible to an observer. After the appropriate interaction between the system and the device, the values of the perceptible macroscopic quantity (P.M.Q.) and the possible values of the system property are correlated in the manner described by the corresponding entangled state. This is often called a pre-measurement. This process is unitary but it is not a measurement. It is the provision of a single result, i.e., a single value of the P.M.Q., which should be referred to as a measurement. Explaining how that happens is the most problematic part which is at the heart of the measurement problem. We have seen in section 2 that the answers provided by the various interpretations are very different (see the taxonomy table). These approaches leave unresolved the explanatory question that motivates the account developed below[28].

The transition from the superposed entangled state to a single perceived outcome is described in ConSol by the hanging-on mechanism. According to this mechanism:

- the observer's perception selects one outcome among the possible branches[29], with probabilities given by the Born rule;
- once selected, the observer's subsequent perceptions remain consistent with this outcome (they "hang on" to the corresponding branch);
- this selection is not a physical process and does not modify the global quantum state.

Thus, the hanging-on mechanism does not introduce any dynamical collapse. It defines a perspectival selection of outcomes, ensuring both definiteness and consistency within a given perspective[30].

## 7.3 Observers as physical systems

This is the key to understanding what happens when you ask another observer which result they obtained after a measurement of a system through an apparatus. In ConSol (as in many perspectival interpretations assuming universality) the other observer is a physical system that you can model by quantum mechanics. This observer plays exactly the same role as the apparatus in the previous description. The observer is like another apparatus that you had coupled to the first one in such a way as to reproduce the results of the first apparatus. The second apparatus would be entangled and correlated with the first one. In this situation, before you interrogate the other observer, the system, the apparatus used in the measurement and the observer who performed the measurement are entangled for you. For you there has been no measurement. Now, you are going to perform a measurement and get a result by interrogating (i.e. measuring) the other observer. So, exactly in the same way that a standard apparatus exhibits a P.M.Q. (for example the position of a needle, a click, or an emission of light from a specific spot) the observer can exhibit a P.M.Q. that we can perceive and that will be linked to the corresponding quantity of the apparatus and the system inside the entanglement chain. This P.M.Q. can be "uttering a sentence" or "making a gesture" for example. Similarly to the fact that a light emission on the top of the screen of a Stern-Gerlach apparatus will be associated with a spin + and a light emission on the bottom of the screen will be associated with a spin -, the utterance of the sentence "I saw +" (or the gesture of raising one hand) will be associated with a spin + and the utterance of the sentence "I saw –" (or the gesture of applauding) will be associated with a spin -. This way of describing seemingly obvious things can seem strange and useless but this is the most important point to understand because the apparent obviousness of this description is precisely what makes it misleading.

Usually when we say "Alice says that she saw +", this sentence has a very high cognitive weight. It means that Alice performed a measurement, that she obtained the result + and that because she obtained the result +, she can utter this sentence expressing precisely what she saw. This seems obvious and of course, that is the usual way we speak. In ConSol we must adopt a different point of view. To understand that, we must come back to what a measurement apparatus is. Recall that it is a device that is built in such a way that when it interacts in an appropriate way with a system, its macroscopic perceptible behaviour can evolve in different ways correlated with the values that a certain property of the system can have. We do not need the details of the physical process leading to the fact that the system can evolve differently in a way that corresponds to the different possible values for the property of the system. The only thing we must assume is that to each different possible value of the property of the system a different P.M.Q. of the apparatus will be associated. Being correlated with the values of the property that is measured on the system which are superposed after the interaction of the apparatus and the system but before the measurement, these P.M.Q.s are entangled with the respective values. Then, observing which P.M.Q. of the apparatus obtains will allow us to say that we have obtained the corresponding value for the property of the system.

[28] I recall that I do not consider in this paper interpretations that modify the standard formalism (dBB theory, GRW theory etc.) which can provide better explanations.

[29] Branches appear due to the selection of a preferred basis by decoherence in the process of measurement. There are no branches independently of a measurement.

[30] If we want to maintain a kind of realist picture of the whole, we can give a possible reason why the perception selects only one of the possible values: this is due to the limits of our perceptual ability which is not able to see the superposition in its fullness. It is similar to the fact that we cannot see infrared or ultraviolet light or we cannot hear sounds of frequency above 40 kHz. Of course we do not claim to be able to give an explicit mechanism of the way our perception works. But it is also possible to get rid of any realist picture and say that the state vector is only a tool allowing us to predict a possible result without representing anything except a set of potentialities for the observer.

Now, it is clear that the fact that an upward-pointing needle is interpreted as a spin + is purely conventional. The important point is that the needle can be either upward-pointing or downward-pointing and that is what allows us to make the distinction between the two possible values of the spin. The opposite convention could have been adopted or even a device where spin + is associated with a right-pointing needle and spin – with a left-pointing needle. Now, we must make the effort to think of the observer in exactly the same way without attaching any special semantic meaning to the answer they give except for the fact that these various answers must be able to be perceived as different. It is easier to think of the two possible answers: raising a hand and applauding (which have no predefined meaning). If these are the answers we can get, without any previous convention we will not know what the meaning of the gesture is. So, if we repeat the experiment many times, we will be able to distinguish between answers meaning that a certain result has been obtained and the others meaning that the other result has been obtained (in case of two possible values). But we will never know which result has been obtained. Let us make the effort to think that it is the same when we hear answers of the form “I saw +”, which, in this picture, we do not have to interpret as “meaning” that the observer actually saw + but rather as one possible value of the P.M.Q. linked to the value of the property of the system. Think of the observer as if they were a physical system reacting in an automatic way exactly as an apparatus. This description concerns only the external modelling of the observer as a physical system. It does not deny that, within their own perspective, the observer has a first-person experience. So you select one of the possible values of the P.M.Q. entangled with the corresponding values of the system. The P.M.Q. is the sound of their voice as they respond. Each possible answer is correlated with a possible result and before you hear one of them all the answers are superposed. Then, by perceiving the sound the observer emits (which is a superposition of all the possible answers), we select one which is the response we hear and which is coupled with the corresponding result. This does not mean that the responding observer has seen this same result as far as they are concerned in their own perspective. The perception of the responding observer is totally outside of this description. In fact, things should be modelled as follows:

1) Point of view of O performing a measurement of the spin along Oz of a particle initially in a state + along Ox: O makes a measurement on the particle using an apparatus. For this, O makes the particle and the apparatus interact appropriately. For O, the particle and the apparatus are then in the entangled state $1/\sqrt{2}$ [|+> |A+> + |-> |A->]. It is assumed that in the state |A+>, the device makes a P.M.Q. appear that is different from the one that appears when it is in the state |A-> (this can be, as we said, the different positions of a needle). O looks at the apparatus and the perception of O then selects one of the possibilities among those offered by the apparatus. The perception of O hangs on to this possibility and O obtains a definite result but nothing physical happens.

2) Point of view of another observer O' about the measurement made by O: O' models the measurement through an entanglement of O with the particle and the apparatus, since O has interacted with both. For O', the observer O, the particle, and the apparatus are in the entangled state 1/√2 [|+> |A+> |O+>+ |-> |A->|O->]. As we emphasized, the states |O+> and |O-> should not be interpreted as indicating that O has "seen" + or "seen" –, but rather as indicating that the values of the P.M.Q. associated with O, if O itself is the object of a measurement, will be different. Just as a position at the top of the screen can be associated with spin + and a position at the bottom with spin -, it can be agreed that the perceptible macroscopic quantity associated with O will be their speech (they will say I saw + or I saw -) or a gesture ("raising the arm" will be correlated with + and "applauding" will be correlated with -). The essential thing is that the P.M.Q. carried by O and associated with the measurement one wants to make can manifest itself in two different values allowing them to be differentiated. O' can then either do the same measurement, i.e. directly look at the apparatus and perceive the P.M.Q. of the device. Or O’ can question O about their result (i.e. measure O) and perceive a P.M.Q. on O (like the sound of the voice or the gesture made by O). The perception of the P.M.Q. (whether that of the device or that of O) will then give rise to a selection of one of the two possible values to which O' will then hang on.

## 7.4 The Tracking assumption in ConSol

Since the hanging-on mechanism applies within each perspective according to the Born rule, without a global rule requiring agreement between perspectives, it is clear that there is no necessary link between the selection made by O' on the P.M.Q. carried by the apparatus (or by O) and the selection made by O on the P.M.Q. of the apparatus. This absence of constraint follows from the fact that the hanging-on mechanism operates independently within each perspective and that no global selection rule exists. The result for O' therefore has no reason to be identical to the result for O. This explains why there is no global perspective. In the case where O' looks at the result directly on the apparatus, the fact that they observe a result possibly different from the one observed by O corresponds to what has been called previously the relativity of the result to the observer. But in the case where O' asks O directly, this corresponds to a violation of the Tracking assumption.

Let us define precisely what the Tracking assumption is in this context. In its usual form, the Tracking assumption presupposes that when one observer asks another observer which result they obtained concerning a certain measurement, what the first observer will hear from the second is precisely what the second obtained. Therefore, Tracking can be formulated as follows:

Let P denote a fixed experimental procedure (for instance, measuring the spin of a particle along a given axis and recording the result). Let Q denote the procedure according to which O asks O' which result they have obtained in the procedure P. Let us denote $R^{(O)}_{O}$ (resp. $R^{(O')}_{O'}$) the result that O (resp. O') obtained in the perspective of O (resp. in the perspective of O') for the measurement P (the spin of the particle). Let us denote $R^{(O)}_{O'\to O}$ the result that O obtains when asking O' which result they obtained (measurement Q). $R^{(O)}_{O}$ and $R^{(O')}_{O'}$ are the results that each observer obtained for the measurement P in their own perspective. $R^{(O)}_{O'\to O}$ is the result that O hears O' reporting when O does the measurement Q. So $R^{(O)}_{O'\to O}$ is the result obtained by O' in O's perspective.

The following equations describe a possible configuration allowed by ConSol under the ideal conditions considered here:

$R^{(O)}_{O} \neq R^{(O')}_{O'}$ (Relativity of outcomes) (13)

$R^{(O)}_{O'\to O} = R^{(O)}_{O}$ (Conviviality) (14)

$R^{(O)}_{O'\to O} \neq R^{(O')}_{O'}$ (Violation of Tracking) (15)

Equations (13) and (15) express possible inequalities, not inequalities that must hold in every case. The outcomes may coincide. The claim is that, for each admissible comparison, a failure of agreement is possible.

In ConSol, relativity of outcomes and violation of Tracking arise from the same underlying mechanism (the hanging-on mechanism), but they are conceptually distinct. Relativity concerns the comparison of results across perspectives, while Tracking concerns the relation between an observer's result in their perspective and another observer's interrogation about that result. Now we can see that modulo Eq. (14) stating Conviviality, Eq. (13) stating Relativity of outcomes is the same equation as Eq. (15) stating Violation of Tracking.

**So it is interesting to notice how ConSol provides a unified view that helps us to understand that the Relativity of outcomes and the Violation of the Tracking assumption are two different forms of the same underlying mechanism, and hence that, under the conviviality condition in Eq. (14), allowing different outcomes across perspectives is incompatible with requiring Tracking to hold universally.**

ConSol combines the possibility of Tracking failure in each admissible instance with strict consistency within each perspective through conviviality.

Comparing the outcomes experienced by different observers requires a meta-level formulation. Such a comparison retains the perspectival indices of the outcomes and does not imply that they coexist as facts within a single observer-independent reality.

[31] We recall that $\Phi^+ = 1/\sqrt{2}\ [|+\rangle|O_+\rangle + |-\rangle\ |O_-\rangle]$.

We see that ConSol fully endorses first-person Universality and is a perfect illustration of Adlam's remark (Adlam, 2024):

> *Ultimately the only thing anyone can ever directly verify is the fact that quantum mechanics makes the right predictions for a single observer.*

We are now also able to clarify this concept of Universality in the context of measurement which was puzzling. As we already noticed in section 6, the way we understand measurement in ConSol shows that Unitarity applies to the description given by an observer of a measurement made by another observer. In this case, the latter observer will get a result but not the first one. The measurement made by Alice is not a measurement for Bob. On the contrary, Unitarity does not apply to the description by an observer of their own measurement.

### 7.5 What happens in a Wigner's friend experiment?

Following what has been presented above we can examine what happens in a Wigner's friend experiment. As long as Wigner has not observed the apparatus inside the laboratory or questioned his friend, the state he correctly attributes is an entangled state in which, for him, no measurement has been made. Indeed, his friend's measurement, which is the fact that his friend's perception has become attached to one of the possible outcomes through the hanging-on mechanism, is not a measurement for him since his perception is not hung on to any value and he has no access to his friend's perceptions. However, the friend can predict that if Wigner asks them a question, Wigner will have a 50% probability of hearing the answer + or – regardless of the result that the friend themselves will have obtained. The friend's prediction regarding the measurement that Wigner may make if he asks is therefore 50% for each result. This is a theoretical reconstruction of Wigner's probabilities within Wigner's perspective. It must be distinguished from the friend's prediction about the report they would themselves hear from Wigner: that report is an outcome within the friend's perspective and is subject to conviviality. But the friend's prediction regarding the measurement that Wigner may make on the whole (friend + system) in the entangled basis containing the projector $|\Phi+\rangle\langle\Phi+|$[31] is that Wigner will always get 1. The friend can therefore put themselves in Wigner's place, knowing that Wigner models them as a quantum system and thus deduce the probability of Wigner's results depending on which measurement he performs. This is reminiscent of what Schmid et al. (2023) say:

> *Consequently, the Friend should not use the state assignments in Eq. (5) or Eq. (6) to make predictions about what Wigner will observe, and rather should assign the state Eq. (4) to herself and system S*[32]*. Consequently, in such relational*

[32] Eqs. (5) and (6) give the states of the friend projected onto one of the possible eigenstates while Eq. (4) is the entangled state.

*views, the Friend agrees with Wigner's prediction. This might seem odd, insofar as the Friend neglects her own information—the specific outcomes they perceived—in making their prediction. But this is exactly what one expects in a perspectival picture where the Friend's outcome is only meaningful relative to the Friend, and is not relevant to Wigner or what he will observe in any sense.*

In principle, this last measurement (in the entangled basis containing the projector |Φ+><Φ+|) can be performed by applying a CNOT gate to the global system and then a Hadamard gate to the system. A measurement on the system and on the friend should then give + and O+ each time. This means that after applying the CNOT gate to the whole and the Hadamard gate to the qubit, if we measure the spin of the qubit and ask the friend, we will get + for the spin and we will hear the friend reply that he saw +. If this result is obtained with certainty after applying the CNOT and Hadamard gates, it confirms the prediction for the entangled state assigned to the composite system before these operations. This is what is predicted by Universality and it is the case in ConSol[33].

ConSol thus reconciles these two apparently conflicting points of view. Contrary to what is stated in the quote above, the friend does not have to neglect their personal information. If they want to predict what Wigner could get as a result, they first have to acknowledge the fact that the actual value that they got is not a result for Wigner himself and second, take into account the type of measurement that Wigner can do. There is no longer any contradiction.

## 7.6 ConSol with respect to the no-go theorem

We now return to the extended Wigner's friend experiment and the assumptions U, AOE, LAOE and SO introduced in Section 5.1.

ConSol fully endorses Universality, in the sense that all physical systems, including observers, can be described by quantum mechanics and evolve unitarily at the level of pre-measurements.

However, ConSol explicitly rejects the assumption of Absoluteness of Observed Events (AOE). In this framework, a measurement outcome is only defined relative to the perspective inside which it is perceived. There is no observer-independent fact specifying the outcome of a measurement. This rejection of AOE is enough to escape the contradiction proved in the no-go theorem.

ConSol distinguishes reversal of the physical measurement interaction from reversal of the experience of obtaining an outcome. The former is allowed in principle. The hanging-on mechanism, however, is not a physical transformation to which an inverse unitary can be applied. This distinction does not by itself establish that ConSol rejects SO in its operational sense. It establishes that reversing the physical interaction must not be identified with undoing the occurrence of a first-person experience.

In ConSol, any application of LAOE must specify the perspective within which the relevant events and measurement settings are described. Cross-perspectival outcome assignments cannot simply be assumed. The response to the no-go theorem developed here rests on rejecting AOE and does not require a separate demonstration about LAOE. Moreover, in ConSol, events within an observer's perspective are constituted by that observer's perceptions and are therefore never spacelike separated. Distant detections reconstructed from perceived records must not be confused with these perceptual events. In an EPR experiment, the relevant correlation is between the observer's own perceived outcome and the answer subsequently heard from the other observer. As explained in Zwirn (2020), the hanging-on mechanism accounts for this correlation without any instantaneous physical change at a distance.

## 7.7 How feral is ConSol?

ConSol rejects an assumption required by the no-go theorem examined here. Now, this comes at the cost of being feral in the most extreme form. I have already explained in a previous paper that among perspectival interpretations (in particular QBism), ConSol is the most perspectival (Zwirn, 2024):

1) The events are relative (a result is a result only for the observer who does the measurement) as is the case in all relativist interpretations. The scenarios themselves are also relative. Events and scenarios are linked to one perspective.

2) No observer can have access to the actual result of another observer (this is the solipsistic part of the interpretation which, actually, is common to all perspectival interpretations that do not respect Tracking). Hence, no observer can legitimately utter a sentence (or make a computation or use joint probabilities) mentioning the result another observer "really" obtained. There is no cross-perspective link. An observer can only mention the answer they heard from asking the other observer which result they obtained. Nevertheless, at the meta-level it is possible (and necessary) to mention the results obtained by each observer within their own perspective.

3) The Tracking assumption is violated. So when an observer hears (in their perspective) the answer of a second observer about the result they obtained, the answer is not necessarily identical to what the second observer obtained in their own perspective.

4) Nevertheless, in the ideal repeatable-measurement scenario considered here, with the relevant correlations and records preserved, an observer who interrogates another observer hears an answer consistent with their own result. This

[33] Recall that we presented this procedure as a way of empirically testing Universality. If the result of the experiment is not always 1, then Universality will have been refuted. The difficulty is that it is not currently possible to apply a CNOT gate to a whole consisting of a system and a friend...

is the convivial aspect of the interpretation.[34]

To summarize, the communication between two observers Alice and Bob follows Eqs. (13) – (15).

As we noticed before, the relativity of events and the violation of the Tracking assumption are two forms of the same underlying mechanism.

Moreover, Alice's scenario and Bob's scenario are not necessarily identical. It is possible that Alice did not perform any measurement in her own scenario.

Being feral in the strongest possible form allows ConSol to escape Riedel's revenge theorems, to which tame relativist interpretations seem to be exposed.

Fortunately there is Conviviality. So, everything happens as though there were a Cross-Perspective link. We cannot witness the ferality. But this is an illusion.

### 7.8 Intersubjectivity and the problem of confirmation

As ConSol is presented as escaping all the issues that other interpretations face, it is necessary to reply to a final criticism raised by Adlam. She worries about the fact that the existence of type III disaccords seems to be a problem of extreme failure of intersubjectivity which undermines the entire practice of science (Adlam, 2022, 2024). If outcomes are defined only relative to individual perspectives, and if the Tracking assumption is violated, it may seem that observers cannot reliably share information about measurement outcomes. This raises a serious concern: how can scientific confirmation remain possible if observers cannot, even in principle, access the results obtained by others?

> *We can straightforwardly accommodate Type-I and Type-II disaccord in a scientific worldview, as long as we don't also have Type-III disaccord - or at least, we don't have Type-III disaccord in ordinary physical interactions like conversations between observers, for it is only that kind of disaccord that would seriously undermine the reliability of the means of communication that we use to establish our shared intersubjective reality.*

So she argues in favour of the possibility of keeping type II disaccord while getting rid of type III disaccord. We are not going to analyse the way she proposes to reach this goal because she herself recognises that due to the Bong et al. theorem it is not possible to achieve that in a way which is compatible with first-person universality of quantum mechanics unless one also rejects another assumption such as Locality or No Superdeterminism[35]. This is the reason why she finally proposes a kind of retrocausality which contradicts No Superdeterminism. Independently of the fact that retrocausality is not very appealing, we will argue that there is no need for that because the danger she sees in type III disaccords for doing science leads to a new, more restrictive way of understanding confirmation but does not undermine the possibility of doing science.

In her article, Adlam (2022) devotes much of the opening discussion to detailing the strange consequences of perspectival interpretations in terms of communication between several observers, principally the fact that the Tracking assumption (that she also calls SF, for Shared Facts) is not respected in this kind of interpretation. As she is not satisfied by this consequence, she insists very clearly on what that means. In particular, she challenges arguments that appear to combine perspectival outcomes with an insufficiently justified assumption of faithful cross-perspective communication. She explicitly says what logically happens in interpretations like QBism and that is not always accepted or even stated clearly by QBists themselves: they have no choice but to accept the fact that the Tracking assumption is not respected. Nevertheless, as I already showed in my review on QBism (Zwirn, 2022) and in another even more detailed paper (Zwirn, 2024), QBists are often reluctant to accept these consequences. Now, QBists have argued in a recent paper (Elia et al., 2025) that intersubjective agreement is unnecessary.

Let us examine the reason why the violation of the Tracking assumption (or SF) means, for her, an impossibility of confirming scientific theories. We must remember that she is a realist aiming to recover an observer-independent reality.

> *It cannot be rational to believe these sorts of interpretations unless they are supplemented with some observer-independent structure which underwrites intersubjective agreement between different observers or at least between successive 'observers' associated with what we would normally take to be the same person.*

She continues by recalling that science is a collaborative endeavour:

> *Typically we put our faith in a theory based not only on our own observations but also the observations that have been made and reported by a multitude of other scientists across the globe. For example, the proponents of the orthodox interpretations clearly have a high level of confidence that quantum mechanics is right, and yet many of them are theoreticians who have not spent much time performing laboratory tests of quantum-mechanical effects, and therefore their own observations certainly do not provide enough data to empirically confirm quantum mechanics as a whole.*

Recall that in the framework of relativist interpretations it is not assumed that there is an external reality that the goal of science is to describe. The only goal of science is to give a consistent account of the personal experience of each observer (in their island universe[36]). Then in perspectival interpretations, what observers can read or hear from others plays exactly the same role as experiments that they could do themselves. Both directly observing an apparatus and reading an experimental report (if the author of the report is reliable) provide perceptual data within the observer's own perspective. Their evidential value depends on the reliability of the relevant experimental and reporting procedures, assessed through evidence available within that perspective. Ultimately, the observer is left with "something" that they can consider as a result (whether it comes from their own experiment or from what they have read about it is not relevant). The experience of an observer comes from everything they can do directly on systems and all that they can read or hear (which are also measurements) about experiments done by others on systems. Science is directed at organising all the private experience of each individual observer. It makes absolutely no difference to assume that this experience comes from real facts shared reliably with other observers or comes from the observer's perceptions that they cannot link to "shared real facts". A difference can be seen only under the assumption that there is an independent reality that plays the role of a referent but it vanishes if there is no such thing. This appears clearly in the following quotation from Adlam:

> *The failure of intersubjectivity in orthodox interpretations of quantum mechanics is significantly more severe [than in classical physics], as even in the ideal case where everything is working perfectly, observers may still disagree. And we have seen that not only will observers disagree in such cases, in general they will not even realise that they disagree.*

[34] The proof is given in many previous papers presenting ConSol.

[35] Actually she said in a recent private communication that she found another way to achieve this goal but I cannot comment here on this new work.

The important point is that she recognises that they cannot realise that they disagree. That is exactly what the convivial feature of ConSol emphasizes. So, the claim that they disagree cannot be made by anybody except at the meta-level. But remember that there is no observer at the meta-level so there is no real observer from whose point of view there is a disagreement. Each real observer's experience is coherent. Science aims to give a consistent account of each personal experience. This is the essential point.

An interesting point is that she denounces the purported solutions proposed by some authors to restore intersubjectivity. This is the case, among others, with Healey (2018), who defends the idea that despite the extreme consequences of these interpretations, one can nevertheless not doubt the sincerity or truthfulness of reports on the results of experiments.

Now there is another issue raised by Adlam who says that two observers doing the same experiments could get different results and so come up with different theories. According to Adlam, a good perspectival interpretation should give an account of the structure of a network of perspectives and not just of a single perspective. She says that science is about the whole network of perspectives and not about the point of view of one perspective. To this objection Rovelli (2024) answers:

> *Imagine that Alice, ideally, does some repeated experiments and finds out a certain specific regularity in the facts relative to herself. She can then communicate to Bob that she has found this regularity. Nothing in RQM questions this possibility. In this case, she is not communicating a contingent fact about the world (these are relational). She is communicating a regularity, Regularities are assumed to be absolute aspects of reality, in RQM, not contingent and relative.*

This does not provide a satisfactory answer to Adlam. Indeed, it is possible to communicate about certain specific regularities. But there is no reason why the communication about regularities will escape the relativist aspect of all communications.

One could argue that the solution to Adlam's concern comes from noticing that confirming for example the Born rule in different perspectives does not require getting exactly the same results, measurement by measurement. All that is required is that the statistics of the results in each perspective follow the Born rule, which would be guaranteed by the hanging-on mechanism inside ConSol even though results can be different when taken one by one.

For Adlam, intersubjective agreement must be grounded in observer-independent facts or in a consistent alignment between different perspectives. ConSol rejects this assumption.

Let us recall that in ConSol, communication between observers involves a physical interaction, while the perception of a definite answer constitutes a measurement for the observer receiving it.

The key point is that this process is subject to the same consistency constraints as any other sequence of outcomes within a perspective. In particular, the conviviality condition ensures that when an observer interrogates another observer about the result of a measurement that they have also performed, the answer they perceive is necessarily consistent with their own previous outcomes.

Adlam's objection is that, in a perspectival framework, one cannot guarantee that the results obtained in perspectives other than one's own also confirm quantum mechanics.

One might argue that the same Born-rule selection mechanism operates in every perspective and therefore guarantees confirmation of quantum mechanics in each of them[37]. This regularity comes from the uniformity of the laws

[36] Adlam quotes Pienaar (2021a) who says that island universes are not motivated by observations and so are ontological excess baggage. But if they are a logical consequence of the way we understand an interpretation, they are not an excess but are mandatory.

[37] The applicability of the Born rule in every perspective must be distinguished from empirical confirmation of that rule in every perspective. ConSol postulates the same Born-rule selection mechanism for each observer, but a probabilistic law allows atypical finite sequences of outcomes. Under the

in all perspectives. Therefore, although different perspectives may contain different outcomes, each perspective confirms quantum mechanics within its own domain of accessible events.

But the above argument, based on a kind of induction about the uniformity of the laws between all perspectives, is not valid. Moreover, a common probabilistic law does not guarantee that every finite sequence of outcomes will empirically confirm that law. We do not even have a single case on which to build the induction since we have never had any access to a perspective other than our own. The question then remains how we can empirically confirm that.

We have to face the facts: we have no way to confirm quantum mechanics in any perspective other than our own.

This leads to a radical shift in epistemological position that I describe in the following:

1. In my perspective I have access to many results coming either from direct experiments that I can perform on quantum systems or from hearing (or reading) reports made by other observers. These reports play exactly the same role as direct experiments.

2. I can confirm science in my own perspective and the theory that I support and which is well confirmed is quantum mechanics.

3. Once I have built this theory which is confirmed inside my perspective, I try to understand what it tells me about the world.

4. It appears that one good way to understand it has consequences about the existence of other perspectives I have no access to. The other perspectives are theoretical consequences of the interpretation of the theory developed to account for the empirical facts internal to my perspective.

5. I have no direct access to the evidence available within other perspectives. An observer whose data are atypical might infer a probabilistic rule different from the Born rule, or even favour a deterministic description. This concerns the conclusions supported by that observer's evidence, not a change in the probabilistic law postulated by ConSol. But whatever happens in another perspective has absolutely no effect inside my own perspective and is totally unknowable.

6. ConSol allows differences between the experienced histories associated with different perspectives, while postulating the same Born-rule selection mechanism for each observer. From within my perspective, I cannot directly inspect those other histories.

7. So the interpretation itself shows that global trans-perspectival confirmation is not a meaningful goal. The very interpretation we are led to, if quantum mechanics is true, shows that the goal of empirically confirming what happens in other perspectives is illusory. This goal is reminiscent of a kind of absoluteness which is precisely what the interpretation shows as false. Confirmation itself is intrinsically perspective-indexed.

8. Now, does that not undermine doing science?

9. Not at all. Despite the fact that a part of the world escapes my access, everything is consistent in the way the interpretation explains the meaning of quantum mechanics.

10. So I can confirm quantum mechanics inside my perspective and see that it leads to a picture where some components are out of reach. Since the theory says that these components are necessary elements of the theory but out of reach, they are outside of the scope of empirical confirmation.

11. I am therefore free to suppose, as an arbitrary assumption, that observers in other perspectives find the same empirical regularities and confirm the same quantum theory as I do within my own perspective, even though I have no means of establishing whether this is actually the case. This supposition is neither empirically verified nor required by the formalism. It is reminiscent of assuming that other humans are conscious despite having no direct access to their experiences.

Scientific confirmation simply has a more limited scope than we might have thought a priori. Zwirn (2000) argued that science already teaches us that some intuitively meaningful goals are in fact inaccessible: undecidable propositions, predictability horizons, or cosmological limits are familiar examples. Perspectival interpretations suggest that global trans-perspectival confirmation belongs to the same category: quantum mechanics endowed with Copenhagenish interpretations shows that the very concept of confirmation is intrinsically perspective-indexed. It remains fully meaningful, but only within the domain where empirical accessibility is defined. What disappears is not scientific rationality, but the classical ideal of a fully observer-independent notion of confirmation.

That does not undermine quantum mechanics because confirming it inside one's perspective leads precisely to the consequence that other perspectives, to which we have no access and that have no effect on our own perspective, exist and that they lie outside the domain where empirical confirmation is meaningful.

This also clarifies why the "ferality" of ConSol is not empirically accessible. Although the theory allows for radical divergence between perspectives at the meta-level, this divergence cannot be experienced by any observer. Each observer is confined to a perspectival trajectory that is internally coherent and in which communication with other observers appears perfectly reliable.

The desire to restore intersubjectivity leads Adlam to propose a cross-perspective link, CPL, as she did in her modification of RQM (Adlam & Rovelli, 2023). She wonders if it is possible to add CPL to other perspectival interpretations and seems to be doubtful. A detailed assessment of RQM, with usual assumptions for repeated independent trials, frequencies close to the Born probabilities become highly probable as the number of trials increases; they are not guaranteed in every finite history. The possibility of atypical histories does not by itself establish that such histories are actually realised. A separate epistemological question is whether an observer can empirically establish what occurs in another perspective. ConSol denies such access.

or without CPL, lies outside the scope of this paper[38]. The question addressed here is whether such a link is necessary for scientific confirmation within a perspective. I find this additional assumption ad hoc and uniquely motivated by the goal of restoring a property that ConSol shows to be unnecessary.

# 8. Conclusion

The analysis conducted throughout this article shows that extended Wigner-type experiments are not merely a technical refinement of the measurement problem, but a challenge to certain assumptions deeply rooted in our classical way of thinking about physical facts. In particular, the logical incompatibility between the universality of unitary dynamics, the absoluteness of observed events, the notion of locality, and the possibility, even in principle, of a super-observer capable of reversing a measurement, forces us to recognise that we cannot simultaneously maintain all of these hypotheses without inconsistency.

Giving up universality amounts to reintroducing a poorly defined boundary between quantum and classical systems, a boundary that we were specifically trying to avoid. Giving up Locality or No Superdeterminism means accepting forms of retrocausality or non-local dependence that greatly disrupt our standard understanding of causality. This is why a growing number of analyses are converging on the idea that it is the absoluteness of observed events that constitutes the main point of tension.

However, saying that events are not absolute is not enough: we must also give a precise and coherent meaning to this non-absoluteness. A simple negation of AOE, left as an interpretive slogan, remains insufficient. What the discussions around type I, II, and III disaccords show, as well as the crucial role played by the Tracking assumption and the cross-perspective link hypothesis, is that the non-absoluteness of events requires rigorous conceptual structuring. Without this, it risks either sinking into ambiguity or implicitly reconstituting itself in the form of a disguised absoluteness.

Many ambiguities in the way perspectival interpretations are often attacked suggest that ordinary language is insufficiently constrained. The appendix introduces a formal system for distinguishing perspective-indexed statements from meta-level combinations of such statements. Its purpose is to make explicit the inferential restrictions adopted here.

In this landscape, Convivial Solipsism appears to offer a particularly stable and conceptually explicit formulation of the measurement process and of what the idea of a "non-absolute event" means. It unambiguously assumes the first-person perspective, clearly distinguishes between the state as a tool for prediction (interaction state) and the observer's lived reality, and rejects any claim of a view from nowhere that would surreptitiously re-establish a global factuality. Events are fully real for the observer who experiences them, without claiming to constitute absolute facts valid independently of any perspective. The meta-level is not a view from nowhere in the absolutist sense, but a descriptive level required to state the rules of the interpretation and to compare indexed perspectival statements. This position preserves the universality of quantum mechanics while giving operational and coherent meaning to the relativity of results.

ConSol addresses Emily Adlam's challenge concerning intersubjectivity and scientific confirmation by locating empirical confirmation within each observer's perspective. The hanging-on mechanism supplies Born-rule probabilities, while conviviality accounts for agreement between an observer's results and the reports they perceive under the ideal conditions specified above. Scientific confirmation does not rely on shared observer-independent facts, but on the stability and consistency of these perspectival structures. Empirical confirmation of other perspectives from within one perspective is an illusory goal.

Convivial Solipsism provides an explicitly perspectival framework in which the meaning of non-absolute events and the inferential restrictions governing their description can be clarified. It shows that the non-absoluteness of events is neither a regression toward idealism nor a negation of scientific objectivity, but rather the expression of a change in the ontological framework: physical reality is no longer conceived as a set of global facts independent of observers, but as a network of coherent descriptions, each anchored in a perspective.

Extended Wigner-type experiments clarify the commitments that interpretations must reconsider if they retain universal unitary dynamics. ConSol develops one response by rejecting absolute outcomes and locating empirical confirmation within perspectives. On this account, scientific objectivity must be understood through the predictive structure and consistency of perspectival experience, rather than through access to observer-independent outcomes.

# 9. Appendix: a Formal Logic for Perspectival Interpretations[39]

## 9.1 The Logic

Here are the basic rules of a formal logic for perspectival interpretations at the first order of relativization. This logic rigorously states the rules to follow when one wants to formulate a sentence inside a perspectival interpretation. The purpose of this appendix is to specify the syntax and inferential rules used in the preceding discussion and to formalize the relation between Tracking and its negation. A full semantic analysis, including the relation between formal possibility and the configurations admitted by ConSol's physical postulates, lies beyond its present scope. This will be developed more fully in a forthcoming paper.

[38] See for example Zwirn (2015) section 3.4.2. or Pienaar's criticisms (Pienaar, 2021b), with which I agree.

### A) Syntax:

The language of this logic consists of the usual logical symbols (∧, ∨, ¬, =>) and the modal symbol ◇, which denotes logical possibility within the formal structure of perspectival descriptions, and the modal symbol □, which denotes logical necessity within the same structure. To these are added the following:

- Type-1 elements: Primary Type-1 elements denoted by $PO^i_j$, which can be interpreted as the fact that observer i has seen result j[40], and compound Type-1 elements obtained recursively by combining primary Type-1 elements with binary propositional logical connectives or by negating any Type-1 element, whether primary or compound.
- Type-2 elements, denoted by $[O_i]$, which are interpreted as indicating the perspective of the observer $O_i$.
- Type-3 elements, denoted as [[Meta]], which are interpreted as indicating the meta-level (a level that is not a perspective in itself but allows us to state, within a single sentence, results observed from different perspectives. The meta-level is descriptive, not absolutist.)

Given the need to always relate a result to a specific perspective, a Type-1 element alone is not a well-formed formula. The corresponding perspective must be specified. Well-formed formulas will be of the form:

1. $[O^i]$ (S) where S is any Type-1 element.

In particular, $[O^i]$ $PO^k_j$ (meaning: "from the perspective of observer $O^i$, observer $O^k$ saw the result j" i.e. if the observer $O^i$ asks the observer $O^k$ which result they obtained, $O^i$ will hear $O^k$ answering "result j" ).

*Remark: ¬ $PO^k_j$ must not be interpreted as "observer $O^k$ saw a result different from result j", but more liberally as "observer $O^k$ saw no result or a result different from result j".*

These formulas will be classified as Type-1 formulas. In brief, a Type-1 formula is a Type-1 element preceded by a Type-2 element.

In the following, let F and G be Type-1 expressions. A primary Type-1 expression is identical to a Type-1 formula. Compound Type-1 expressions are constructed recursively by combining Type-1 expressions using binary propositional logical connectives, by negating a Type-1 expression, or by prefixing a Type-1 expression with a modal operator, □ or ◊. These operations may be applied to primary or compound Type-1 expressions. Compound Type-1 expressions are not well-formed formulas and cannot be used without the prefix [[Meta]].

Let H and K be Type-1 elements. Let @ stand for a binary propositional logical connective.

The following formulas are well-formed and will be classified as Type-2 formulas.

2. [[Meta]] F (primary Type-2 formula)
3. Compound Type-2 formula obtained recursively by combining primary Type-2 formulas through binary propositional logical connectives, negation and modal operators.

Logical combinations at the meta-level concern indexed perspectival statements and must not be interpreted as defining a single observer-independent state of affairs.

*Important point: Notice that no formula of the form [[Meta]] $PO^k_j$ is well-formed. The meta-level does not remove perspectival indexing. $[O^i]$ [[Meta]] F is not allowed either.*

### B) Axioms:

A1. [[Meta]] (¬ (F∧ ¬ F))

A2. ¬ {([[Meta]] (¬ F)) ∧ ([[Meta]] (F))}

A3. $[O^i]$ (¬ (H∧ ¬ H))

A4. [[Meta]] (¬ {($[O^i]$ (¬ H)) ∧ ($[O^i]$ (H))})

These axioms express consistency requirements at the perspectival level and at the meta-level. They allow one to derive a logical contradiction from a contradiction in a formula following $[O^i]$ or [[Meta]].

### C) Deduction rules:

R1. $[O^i]$ H ⊢ [[Meta]] $[O^i]$ H

*Remark: note that one cannot write F ⊢ [[Meta]] F because, except when F consists of a single Type-1 formula, F is not a well-formed formula.*

R2. [[Meta]] $[O^i]$ H ⊢ $[O^i]$ H

*Remark: note that one cannot write [[Meta]] F ⊢ F because, except when F is reduced to one single Type-1 formula, F is not a well-formed formula.*

R3. [[Meta]] (¬F) ⊢ ¬ [[Meta]] (F)

R4. ¬ [[Meta]] (F) ⊢ [[Meta]] (¬F)

R5. [[Meta]] (F @ G) ⊢ ([[Meta]] F) @ ([[Meta]] G)

R6. ([[Meta]] F) @ ([[Meta]] G) ⊢ [[Meta]] (F @ G)

R7. [[Meta]] {$[O^i]$ (¬H)} ⊢ [[Meta]] {¬ $[O^i]$ (H)}

R8. [[Meta]] {¬ $[O^i]$ (H)} ⊢ [[Meta]] {$[O^i]$ (¬H)}

[40] The reason why we adopt the notation $PO^i_j$ is that it can be understood as "Perception of $O^i$ is result j".

R9. $[O^i]$ (H @ K) ⊢ [[Meta]] (($[O^i]$ H) @ ($[O^i]$ K))

R10. [[Meta]] (($[O^i]$ H) @ ($[O^i]$ K)) ⊢ $[O^i]$ (H @ K)

R11. [[Meta]] □ F ≡ □ [[Meta]] F (≡ is the symbol of demonstrable equivalence)

R12. [[Meta]] ◇ F ≡ ◇ [[Meta]] F

R13. The rules of classical propositional logic and of modal logic K restricted to the allowed categories of formulas.

### 9.2 Examples of application

The Tracking assumption can be formulated as:

TR. [[Meta]] (□ {$[O^i]$ $PO^k_j$ => $[O^k]$ $PO^k_j$}) for i≠k

While the violation of Tracking is:

VTR. [[Meta]] (◇ {$[O^i]$ $PO^k_j$ ∧ $[O^k]$ (¬$PO^k_j$ )}) for i≠k

Proof: the negation of TR is (we omit the condition for i≠k in what follows):

¬ [[Meta]] (□ {$[O^i]$ $PO^k_j$ => $[O^k]$ $PO^k_j$})

Which gives by R4:

[[Meta]] ¬ (□ {$[O^i]$ $PO^k_j$ => $[O^k]$ $PO^k_j$})

Which is:

[[Meta]] (◇ {$[O^i]$ $PO^k_j$ ∧ ¬ $[O^k]$ ($PO^k_j$ )})

And by R8:

[[Meta]] {¬ $[O^k]$ ($PO^k_j$)} ⊢ [[Meta]] {$[O^k]$ (¬$PO^k_j$ )}

And by R7:

[[Meta]] {$[O^k]$ (¬$PO^k$j)} ⊢ [[Meta]]{¬ $[O^k]$ ($PO^k_j$ )}

Hence:

[[Meta]] {$[O^k]$ (¬$PO^k$j)} ≡ [[Meta]]{¬ $[O^k]$ ($PO^k_j$ )}

Now by R12:

[[Meta]] (◇ {$[O^i]$ $PO^k_j$ ∧ $[O^k]$ (¬$PO^k_j$ )}) ≡ ◇ [[Meta]] ({$[O^i]$ $PO^k_j$ ∧ $[O^k]$ (¬$PO^k_j$ )})

Since [[Meta]] {$[O^k]$ (¬$PO^k$j)} ≡ [[Meta]]{¬ $[O^k]$ ($PO^k_j$ )}, we get by R5, R6 and the replacement of equivalents proven in K-logic:

◇ [[Meta]] ({$[O^i]$ $PO^k_j$ ∧ $[O^k]$ (¬$PO^k_j$ )})≡ ◇ [[Meta]]{ $[O^i]$ $PO^k_j$ ∧ ¬ $[O^k]$ ($PO^k_j$ )}

And by R12 again:

◇ [[Meta]]{ $[O^i]$ $PO^k_j$ ∧ ¬ $[O^k]$ ($PO^k_j$ )}≡ [[Meta]] (◇{$[O^i]$ $PO^k_j$ ∧¬ $[O^k]$ ($PO^k_j$ )}).

ConSol follows the rules of this logic with the axiom of violation of Tracking.

In the majority of attempts that mainly use ordinary, non-formalized sentences to prove a contradiction inside ConSol, frequent errors come from forgetting that $PO^i_j$ alone is not a well-formed formula or from combining perspectival propositions such as $[O^i]$ $PO^k_i$ with $[O^k]$ $PO^k_j$ outside the meta-level. They can be combined only at the meta-level (i.e. [[Meta]] ($[O^i]$ $PO^k_i$ ∧ $[O^k]$ $PO^k_j$)).

**Funding:** This research received no external funding.

**Acknowledgements:** This paper has benefited from many exchanges with Emily Adlam, Eric Cavalcanti, Dennis Dieks, Markus Muller, Tim Riedel and Jean Sass. I would also like to thank Jeffrey Barrett, Roger Balian, Michel Bitbol, Caslav Brukner, Andrea Di Biagio, Chris Fuchs, Philippe Grangier, Richard Healey, Franck Laloë, Bruno Mansoulié, and Alex Matzkin for many enlightening discussions and for having read some earlier drafts of this paper.

**Conflict of interest:** The author declares no conflict of interest. The opinions expressed in this paper are those of the author and do not reflect the views of the people quoted in the acknowledgements.

## References

Adlam, E. (2022). Does science need intersubjectivity? The problem of confirmation in orthodox interpretations of quantum mechanics. *Synthese, 200,* Article 522. https://doi.org/10.1007/s11229-022-03989-0

Adlam, E. (2024). What does '(non)-absoluteness of observed events' mean? *Foundations of Physics, 54,* Article 13. https://doi.org/10.1007/s10701-023-00747-1

Adlam, E., & Rovelli, C. (2023). Information is physical: Cross-perspective links in relational quantum mechanics. *Philosophy of Physics, 1*(1), Article 4. https://doi.org/10.31389/pop.8

Albert, D. Z., & Loewer, B. (1988). Interpreting the many worlds interpretation. *Synthese, 77*(2), 195-213. https://doi.org/10.1007/BF00869434

Auffèves, A., & Grangier, P. (2016). Contexts, systems and modalities: A new ontology for quantum mechanics. *Foundations of Physics, 46*, 121-137. https://doi.org/10.1007/s10701-015-9952-z

Barrett, J. A. (1999). *The quantum mechanics of minds and worlds*. Oxford University Press.

Bene, G., & Dieks, D. (2002). A perspectival version of the modal interpretation of quantum mechanics and the origin of macroscopic behavior. *Foundations of Physics, 32*, 645-671. https://doi.org/10.1023/A:1016014008418

Bohm, D. (1952a). A suggested interpretation of the quantum theory in terms of "hidden" variables. I. *Physical Review, 85*, 166-179. https://doi.org/10.1103/PhysRev.85.166

Bohm, D. (1952b). A suggested interpretation of the quantum theory in terms of "hidden" variables. II. *Physical Review, 85*, 180-193. https://doi.org/10.1103/PhysRev.85.180

Bong, K.-W., Utreras-Alarcón, A., Ghafari, F., Liang, Y.-C., Tischler, N., Cavalcanti, E. G., Pryde, G. J., & Wiseman, H. M. (2020). A strong no-go theorem on the Wigner's friend paradox. *Nature Physics, 16*, 1199-1205. https://doi.org/10.1038/s41567-020-0990-x

Brassard, G., & Raymond-Robichaud, P. (2020). *Parallel lives: A local-realistic interpretation of "nonlocal" boxes* (Version 3). arXiv. https://arxiv.org/abs/1709.10016v3

Brukner, Č. (2017). On the quantum measurement problem. In R. Bertlmann & A. Zeilinger (Eds.), *Quantum [Un]Speakables II: Half a century of Bell's theorem* (pp. 95-117). Springer. https://doi.org/10.1007/978-3-319-38987-5_5

Brukner, Č. (2018). A no-go theorem for observer-independent facts. *Entropy, 20*(5), Article 350. https://doi.org/10.3390/e20050350

Cohen-Tannoudji, C., Diu, B., & Laloë, F. (2019). *Quantum mechanics*. Wiley.

d'Espagnat, B. (1971). *Conceptual foundations of quantum mechanics*. W. A. Benjamin.

DeWitt, B. S., & Graham, N. (Eds.). (1973). *The many-worlds interpretation of quantum mechanics*. Princeton University Press.

Di Biagio, A., & Rovelli, C. (2021). Stable facts, relative facts. *Foundations of Physics, 51*, Article 30. https://doi.org/10.1007/s10701-021-00429-w

Di Biagio, A., & Rovelli, C. (2025). *Relative information, relative facts*. arXiv. https://arxiv.org/abs/2510.11349

Dieks, D. (2022). Perspectival quantum realism. *Foundations of Physics, 52*, Article 95. https://doi.org/10.1007/s10701-022-00611-8

Dieks, D. (2025). Radical perspectivalism, locality and relativity. *Foundations of Physics, 55*, Article 30. https://doi.org/10.1007/s10701-025-00844-3

Diósi, L. (1989). Models for universal reduction of macroscopic quantum fluctuations. *Physical Review A, 40*, 1165-1174. https://doi.org/10.1103/PhysRevA.40.1165

Elia, G., Carter, J., & Crease, R. (2025). *Intersubjective agreement about quantum states is unnecessary in QBism* (Version 1). arXiv. https://arxiv.org/abs/2508.16683v1

Everett, H. (1957a). *On the foundations of quantum mechanics* [Doctoral dissertation, Princeton University].

Everett, H. (1957b). "Relative state" formulation of quantum mechanics. *Reviews of Modern Physics, 29*, 454-462. https://doi.org/10.1103/RevModPhys.29.454

Faye, J. (2019). Copenhagen interpretation of quantum mechanics. In E. N. Zalta (Ed.), *The Stanford encyclopedia of philosophy* (Winter 2019 ed.). Metaphysics Research Lab, Stanford University. https://plato.stanford.edu/archives/win2019/entries/qm-copenhagen/

Fuchs, C. A., & Schack, R. (2013). Quantum-Bayesian coherence. *Reviews of Modern Physics, 85*, 1693-1715. https://doi.org/10.1103/RevModPhys.85.1693

Fuchs, C. A., & Schack, R. (2014). *Quantum measurement and the Paulian idea*. arXiv. https://arxiv.org/abs/1412.4209

Fuchs, C. A., & Schack, R. (2015). QBism and the Greeks: Why a quantum state does not represent an element of physical reality. *Physica Scripta, 90*(1), Article 015104. https://doi.org/10.1088/0031-8949/90/1/015104

Fuchs, C. A., & Stacey, B. C. (2019). QBism: Quantum theory as a hero's handbook. In E. M. Rasel, W. P. Schleich, & S. Wölk (Eds.), *Foundations of quantum theory* (Proceedings of the International School of Physics "Enrico Fermi", Vol. 197, pp. 133-202). IOS Press; Società Italiana di Fisica. https://doi.org/10.3254/978-1-61499-937-9-133

Ghirardi, G. C., Rimini, A., & Weber, T. (1986). Unified dynamics for microscopic and macroscopic systems. *Physical Review D, 34*, 470-491. https://doi.org/10.1103/PhysRevD.34.470

Harrigan, N., & Spekkens, R. W. (2010). Einstein, incompleteness, and the epistemic view of quantum states. *Foundations of Physics, 40*, 125-157. https://doi.org/10.1007/s10701-009-9347-0

Healey, R. (2012). Quantum theory: A pragmatist approach. *The British Journal for the Philosophy of Science, 63*, 729-771. https://doi.org/10.1093/bjps/axr054

Healey, R. (2016). Quantum-Bayesian and pragmatist views of quantum theory. In E. N. Zalta (Ed.), *The Stanford encyclopedia of philosophy* (Winter 2016 ed.). Metaphysics Research Lab, Stanford University. https://plato.stanford.edu/archives/win2016/entries/quantum-bayesian/

Healey, R. (2017). *The quantum revolution in philosophy*. Oxford University Press.

Healey, R. (2018). Quantum theory and the limits of objectivity. *Foundations of Physics, 48*, 1568-1589. https://doi.org/10.1007/s10701-018-0216-6

Healey, R. (2025). Quantum theory from a pragmatist perspective. In H. Zwirn (Ed.), *Quantum physics and cosmology: The mysteries of the infinitely small and the infinitely large* (pp. 79-108). ISTE; Wiley.

Kochen, S., & Specker, E. (1967). The problem of hidden variables in quantum mechanics. *Journal of Mathematics and Mechanics, 17*(1), 59-87. https://doi.org/10.1512/iumj.1968.17.17004

Laudisa, F., & Rovelli, C. (2021). Relational quantum mechanics. In E. N. Zalta (Ed.), *The Stanford encyclopedia of philosophy* (Spring 2021 ed.). Metaphysics Research Lab, Stanford University. https://plato.stanford.edu/archives/spr2021/entries/qm-relational/

Leifer, M. S. (2014). Is the quantum state real? An extended review of ψ-ontology theorems. *Quanta, 3*, 67-155. https://doi.org/10.12743/quanta.v3i1.22

Okon, E. (2025). Reassessing the strength of a class of Wigner's friend no-go theorems. *Entropy, 27*(6), Article 563. https://doi.org/10.3390/e27060563

Ormrod, N., & Barrett, J. (2022). *A no-go theorem for absolute observed events without inequalities or modal logic*. arXiv. https://arxiv.org/abs/2209.03940

Penrose, R. (1996). On gravity's role in quantum state reduction. *General Relativity and Gravitation, 28*, 581-600. https://doi.org/10.1007/BF02105068

Pienaar, J. (2021a). QBism and relational quantum mechanics compared. *Foundations of Physics, 51*, Article 96. https://doi.org/10.1007/s10701-021-00501-5

Pienaar, J. (2021b). A quintet of quandaries: Five no-go theorems for relational quantum mechanics. *Foundations of Physics, 51*, Article 97. https://doi.org/10.1007/s10701-021-00500-6

Pienaar, J. (2024). *A single space-time is too small for all of Wigner's friends* (Version 2). arXiv. https://arxiv.org/abs/2312.11759v2

Riedel, T. (2024). Relational quantum mechanics, quantum relativism, and the iteration of relativity. *Studies in History and Philosophy of Science, 104*, 109-118. https://doi.org/10.1016/j.shpsa.2024.02.007

Riedel, T. (in press). Is quantum relativism untameable? Revenge Wigner arguments for relative facts. *The British Journal for the Philosophy of Science*. https://doi.org/10.1086/732830

Rovelli, C. (1996). Relational quantum mechanics. *International Journal of Theoretical Physics, 35*(8), 1637-1678. https://doi.org/10.1007/BF02302261

Rovelli, C. (2024). *Can Alice do science and have friends, in a relational quantum world? Solipsism and relational quantum mechanics*. arXiv. https://arxiv.org/abs/2410.20012

Schmid, D., Yīng, Y., & Leifer, M. S. (2023). *A review and analysis of six extended Wigner's friend arguments*. arXiv. https://arxiv.org/abs/2308.16220

Schmid, D., Yīng, Y., & Leifer, M. S. (2025). Copenhagenish interpretations of quantum mechanics. In H. Zwirn (Ed.), *Quantum physics and cosmology: The mysteries of the infinitely small and the infinitely large* (pp. 133-170). ISTE; Wiley. https://doi.org/10.1002/9781394427673.ch7

Spekkens, R. W. (2007). Evidence for the epistemic view of quantum states: A toy theory. *Physical Review A, 75*, Article 032110. https://doi.org/10.1103/PhysRevA.75.032110

Van Den Bossche, M., & Grangier, P. (2023). *Revisiting quantum contextuality in an algebraic framework*. arXiv. https://arxiv.org/abs/2304.07757

Wigner, E. P. (1962). Remarks on the mind-body question. In I. J. Good (Ed.), *The scientist speculates* (pp. 284-301). Basic Books.

Wiseman, H. M., Cavalcanti, E. G., & Rieffel, E. G. (2023). A "thoughtful" Local Friendliness no-go theorem: A prospective experiment with new assumptions to suit. *Quantum, 7*, Article 1112. https://doi.org/10.22331/q-2023-09-14-1112

Zwirn, H. (2000). *Les limites de la connaissance*. Odile Jacob.

Zwirn, H. (2015). *The measurement problem: Decoherence and Convivial Solipsism* [Long version]. arXiv. https://arxiv.org/abs/1505.05029

Zwirn, H. (2016). The measurement problem: Decoherence and Convivial Solipsism. *Foundations of Physics, 46*, 635-667. https://doi.org/10.1007/s10701-016-9999-5

Zwirn, H. (2020). Nonlocality versus modified realism. *Foundations of Physics, 50*, 1-26. https://doi.org/10.1007/s10701-019-00314-7

Zwirn, H. (2021). Is the past determined? *Foundations of Physics, 51*, Article 57. https://doi.org/10.1007/s10701-021-00459-4

Zwirn, H. (2022). Is QBism a possible solution to the conceptual problems of quantum mechanics? In O. Freire Jr., G. Bacciagaluppi, O. Darrigol, T. Hartz, C. Joas, A. Kojevnikov, & O. Pessoa Jr. (Eds.), *The Oxford handbook of the history of quantum interpretations* (pp. 1007-1030). Oxford University Press. https://doi.org/10.1093/oxfordhb/9780198844495.013.0042

Zwirn, H. (2023). Everett's interpretation and Convivial Solipsism. *Quantum Reports, 5*, 267-281. https://doi.org/10.3390/quantum5010018

Zwirn, H. (2024). Convivial Solipsism as a maximally perspectival interpretation. *Foundations of Physics, 54*, Article 39. https://doi.org/10.1007/s10701-024-00773-7

Zwirn, H. (2025). The role of the observer: Convivial Solipsism. In H. Zwirn (Ed.), *Quantum physics and cosmology: The mysteries of the infinitely small and the infinitely large* (pp. 109-132). ISTE; Wiley.

Zwirn, H. (2026). Are events absolute? *EPJ Web of Conferences, 366*, Article 01002. https://doi.org/10.1051/epjconf/202636601002